\documentclass[pra,nofootinbib,floats,aps,twocolumn,tightenlines,superscriptaddress]{revtex4-1}

\usepackage{graphicx}
\usepackage{bbold}
\usepackage{mathtools}
\usepackage{hyperref}
\usepackage{commath}
\usepackage{bm}
\usepackage{amsfonts,amsmath,amssymb,amsthm}
\usepackage[usenames,dvipsnames]{xcolor}
\usepackage{subfigure}
\usepackage{dsfont}

\usepackage{simplewick}

\DeclareMathOperator{\Tr}{Tr}

\renewcommand*\d[2][]{%
	\mathrm{d}%
	\ifx\relax#1\relax\else
	\rule{-0.02em}{1.5ex}^{#1}\rule{0.08em}{0ex}\!
	\fi
	#2\,
}
\newcommand{\ket}[1]{| {#1} \rangle}
\newcommand{\bra}[1]{\langle {#1} |}
\newcommand{\tr}{\text{Tr}}

\newcommand{\ii}{\mathrm{i}}

\renewcommand{\a}[1]{\hat{a}_{\bm{#1}}}
\newcommand{\ad}[1]{\hat{a}_{\bm{#1}}^\dagger}
\newcommand{\an}[1]{\hat{a}_{#1}}
\newcommand{\adn}[1]{\hat{a}_{#1}^\dagger}
\newcommand{\ah}{\hat{a}}
\newcommand{\rhoa}{\hat{\rho}_\textsc{a}}
\newcommand{\rhoab}{\hat{\rho}_\textsc{ab}}
\newcommand{\rhoabpt}{\hat{\rho}_\textsc{ab}^{{\text{\textbf{t}}}_\textsc{a}}}
\newcommand{\ma}{\hat{m}_\textsc{a}}
\newcommand{\mb}{\hat{m}_\textsc{b}}
\newcommand{\mua}{\hat{\mu}_\textsc{a}}
\newcommand{\mub}{\hat{\mu}_\textsc{b}}
\newcommand{\Ya}{\hat{Y}_\textsc{a}}
\newcommand{\Yb}{\hat{Y}_\textsc{b}}

\newcommand{\gak}{\ket{\tilde{g}_\textsc{a}}}
\newcommand{\eak}{\ket{\tilde{e}_\textsc{a}}}
\newcommand{\gbk}{\ket{\tilde{g}_\textsc{b}}}
\newcommand{\ebk}{\ket{\tilde{e}_\textsc{b}}}
\newcommand{\gab}{\bra{\tilde{g}_\textsc{a}}}
\newcommand{\eab}{\bra{\tilde{e}_\textsc{a}}}
\newcommand{\gbb}{\bra{\tilde{g}_\textsc{b}}}

\newcommand{\alphak}{\ket{\alpha(\bm k)}}
\newcommand{\alphab}{\bra{\alpha(\bm k)}}
\newcommand{\disp}{\hat{D}_{\alpha(\bm{k})}}
\newcommand{\ca}{\mathcal{C}_\textsc{a}}
\newcommand{\cb}{\mathcal{C}_\textsc{b}}
\newcommand{\cnu}{\mathcal{C}_\nu}
\newcommand{\cmu}{\mathcal{C}_\mu}
\newcommand{\fa}{f_\textsc{a}}
\newcommand{\fb}{f_\textsc{b}}
\newcommand{\fp}{f_\text{p}}
\newcommand{\fm}{f_\text{m}}
\newcommand{\bat}{\tilde{\beta}_\textsc{a}(\bm k)}
\newcommand{\bbt}{\tilde{\beta}_\textsc{b}(\bm k)}
\newcommand{\intk}{\begingroup\textstyle \int\endgroup \!\d[n]{\bm k}}
\newcommand{\normord}[1]{:\mathrel{#1}:}
\newcommand{\ba}{\beta_\textsc{a}(\bm k)}
\newcommand{\bb}{\beta_\textsc{b}(\bm k)}
\newcommand{\bnu}{\beta_\nu(\bm k)}

\begin{document}

\title{Non-perturbative analysis of entanglement harvesting from coherent field states}
	
\author{Petar Simidzija}	\affiliation{Department of Applied Mathematics, University of Waterloo, Waterloo, Ontario, N2L 3G1, Canada}
\affiliation{Institute for Quantum Computing, University of Waterloo, Waterloo, Ontario, N2L 3G1, Canada}
	
\author{Eduardo Mart\'in-Mart\'inez}
\affiliation{Department of Applied Mathematics, University of Waterloo, Waterloo, Ontario, N2L 3G1, Canada}
\affiliation{Institute for Quantum Computing, University of Waterloo, Waterloo, Ontario, N2L 3G1, Canada}
\affiliation{Perimeter Institute for Theoretical Physics, Waterloo, Ontario N2L 2Y5, Canada}

\begin{abstract}
We study (non-perturbatively) the interactions of delta-coupled detectors with coherent states of a scalar field. We show that the time-evolved density matrix spectra of i) a single detector, ii) two detectors, and iii) the partial transpose of the latter, are all independent of which coherent state the field was in. Furthermore, we find that the eigenvalues in iii) are non-negative, implying that a delta-coupled detector pair cannot harvest entanglement from the vacuum, or any other coherent field state.
\end{abstract}
	
\maketitle
	
\section{Introduction}
\label{sec:intro}

It is a well-known result of quantum field theory that different regions of a quantum field vacuum contain correlations (including  entanglement), even if the regions are spacelike separated~\cite{Summers1985,Summers1987}. This finding has been important not only to our understanding of long-standing open problems, such as the black hole information loss problem~\cite{Preskill1992}, but also in the emergence of novel and interesting phenomena, such as, for instance, Masahiro Hotta's \textit{quantum energy teleportation}~\cite{Hotta2008,Hotta2009,Frey2014}. 

Another phenomenon that makes use of the entanglement present in a quantum field has come to be known as \textit{entanglement harvesting}. The pioneering works by Valentini~\cite{Valentini1991}, and later by Reznik \textit{et. al.}~\cite{Reznik2003,Reznik2005}, showed that entanglement from a scalar field vacuum can be operationally extracted by a pair of initially unentangled quantum two-level  particle detectors.

Following these initial studies, there has been a significant amount of work done to understand how sensitive entanglement harvesting is on the various components of its setup. This has been largely motivated by the realization that entanglement is of critical importance in the field of quantum information processing (see e.g.~\cite{Nielsen2010}). For instance, on the more fundamental side, it has been shown that entanglement harvesting is sensitive to the geometry~\cite{Steeg2009} and topology~\cite{Martinez2016a} of the background spacetime.

From the perspective of working towards an experimental implementation of an entanglement harvesting protocol, it has been found that harvestable entanglement is not a unique feature of the scalar field vacuum. For instance, the ability of an Unruh-DeWitt (UDW) particle detector pair to harvest entanglement from a scalar field vacuum (by coupling to either the field's amplitude or momentum) is qualitatively similar to the ability of fully-featured hydrogenlike atoms to harvest entanglement from the electromagnetic field vacuum~\cite{Pozas2016}. In fact, it has been suggested that space- and timelike entanglement harvesting protocols are experimentally feasible with current tools in atomic and superconducting systems~\cite{Olson2011,Olson2012,Sabin2012}.

%Before an experimental realization of an entanglement harvesting protocol comes to fruition, it is important that we fully understand how best to optimize the amount of harvestable entanglement. Perhaps the two most straightforward ways to do this are to i) vary the properties of the particle detectors, or ii) alter the state of the field. For instance, it has been shown that a non-zero detector energy gap is necessary for entanglement harvesting to be possible. Meanwhile as an example of ii), it was shown that, perhaps unsurprisingly, the amount of entanglement that can be harvested from thermal scalar field states decays rapidly with increasing temperature~\cite{Brown2013a}.

Here, we are interested in how sensitive the entanglement harvesting ability of an UDW detector pair is to the detectors being shone by coherent light (more precisely, how sensitive it is to which coherent scalar field state the detectors interact with). In~\cite{Simidzija2017b} it was proven that to second order in the detectors' coupling strengths, the evolved one- and two-detector density matrix eigenvalues, as well as the eigenvalues of the partially-transposed two-detector density matrix, are independent of which coherent field state the detector/detectors interacted with. Therefore, perturbatively, a detector pair can harvest the same amount of entanglement from any coherent field state. 

In this paper we will show that these rather remarkable results also hold in the non-perturbative regime, at least for detectors that interact with the field through a short and intense coupling (formally described by a Dirac-delta switching function). Furthermore, we will extend to the non-perturbative regime one of the perturbative results in~\cite{Pozas2017}. Namely, in~\cite{Pozas2017}, it was shown that a delta coupled detector cannot harvest entanglement from the field vacuum to leading order in perturbation theory. Here we will show non-perturbatively that delta-coupled detectors (with arbitrary spatial profiles, field coupling strengths and energy gaps) \textit{cannot} harvest entanglement from the vacuum, or any other coherent state of the scalar field at all.

After defining coherent states of a scalar field in Sec.~\ref{sec:field}, and introducing the formalism of the Unruh-DeWitt detector-field interaction model in Sec.~\ref{sec:UDW}, we organize the main claims of this paper into three main sections. First, in Sec.~\ref{sec:one_detector}, we prove that a single detector's evolved density matrix spectrum is independent of which coherent field state it interacted with. We then show, in Sec.~\ref{sec:two_detectors}, that the evolved two-detector density matrix spectrum, as well as the spectrum of the partially transposed two-detector density matrix, are also independent of the field's coherent amplitude. In Sec.~\ref{sec:ent_harv}, we prove that the eigenvalues of the partially transposed matrix are always non-negative, and therefore that a detector pair cannot harvest entanglement from any coherent field state. We make our conclusions in Sec.~\ref{sec:conclusions}, and additional technical details are provided in the Appendices. Natural units $\hbar=c=1$ are used throughout.

\section{Setup}
\label{sec:setup}

\subsection{Coherent states of a scalar field}
\label{sec:field}

We consider a scalar quantum field $\hat{\phi}(\bm{x},t)$ in $(n+1)$-dimensional, flat spacetime. Written in terms of plane wave modes, $\hat{\phi}(\bm{x},t)$ takes the form
\begin{equation}
\label{eq:field}
	\hat{\phi}(\bm{x},t)
	=
	\int\frac{\d[n]{\bm{k}}}{\sqrt{2(2\pi)^n |\bm{k}|}}\left[\ad{k} e^{\ii(|\bm{k}|t-\bm{k}\cdot\bm{x})}+\text{H.c.}\right].
\end{equation}
Here, $\ad{k}$ and $\a{k}$ are bosonic creation and annihilation operators, respectively. The lowest energy state (i.e. the vacuum state) of the free field is denoted $\ket{0}$, and it satisfies the condition
\begin{equation}
\label{eq:ground_state}
	\a{k}\ket{0}=0,
\end{equation}
for all $\bm{k}\in {\rm I\!R}^n$. 

We now define a general coherent state of the field $\ket{\alpha(\bm k)}$ to be a displaced vacuum state:
\begin{equation}
\label{eq:alpha}
	\ket{\alpha(\bm{k})}
	\coloneqq
	\hat{D}_{\alpha(\bm{k})}\ket{0}
	\coloneqq
	\exp\!
	\left(\int\!\!\d[n]{\bm{k}}\!\!\left[\alpha(\bm{k})\ad{k}
	\!-\!
	\alpha^*(\bm{k})\a{k}\right]\!\right)
	\!\!\ket{0},
\end{equation}
where the unitary operator $\hat{D}_{\alpha(\bm{k})}$ is often referred to as a \textit{displacement operator}, since it implements translations in phase space. The complex valued distribution $\alpha(\bm k)$, which we call the \textit{coherent amplitude}, characterizes the coherent state $\ket{\alpha(\bm k)}$. In quantum optics, coherent states of the electromagnetic field are commonly used to describe coherent light~\cite{scully1999}, in which the photon number is Poisson distributed. In this setting, we find a physical interpretation for the coherent amplitude: namely, $|\alpha(\bm k)|^2$ is a distribution of the average number of photons of wavenumber $\bm k$ in the state $\alphak$. Although we are considering a scalar field instead of the electromagnetic field, we will, in a qualitative manner, refer back to this physical interpretation of coherent states as describing coherent light.

Note that coherent states are the eigenstates of the $\a{k}$ operators (acting from the left):
\begin{equation}
\label{eq:coh_1}
	\a{k'}\ket{\alpha(\bm{k})}=\alpha(\bm{k'})\ket{\alpha(\bm{k})}.
\end{equation}
Let us show that our definition~\eqref{eq:alpha} of a coherent state $\ket{\alpha(\bm{k})}$ satisfies~\eqref{eq:coh_1}. Using the Baker-Campbell-Hausdorff formula (see, among others,~\cite{Truax1988}) together with the canonical commutation relations, we find that
\begin{equation}
\label{eq:commutator}
	\left[
	\a{k'},\hat{D}_{\alpha(\bm{k})}
	\right]=
	\alpha(\bm{k}')\hat{D}_{\alpha(\bm{k})}.
\end{equation}
Multiplying on the right by $\ket{0}$ and using the condition $\a{k'}\ket{0}=0$ results in~\eqref{eq:coh_1}.

\subsection{Detector(s)-field evolution}
\label{sec:UDW}

\subsubsection{One detector}

We first consider a single particle  detector (labelled A) interacting with the field. We denote the initial time density matrix of the detector-field system by $\hat{\rho}_\text{1,i}$, where the subscript 1 indicates that there is one detector. We will describe the time evolution of the system using the well-known Unruh-DeWitt model~\cite{DeWitt1979}, which successfully captures most of the relevant features of the light-matter interaction when angular momentum exchange can be ignored~\cite{Martinez2013,Alhambra2014,Pozas2016}. In this model, the particle detector is a two-level quantum system with a free Hamiltonian ground state $\ket{g_\textsc{a}}$, excited state $\ket{e_\textsc{a}}$, and energy gap $\Omega_\textsc{a}$. We assume the detector to be at rest in an inertial reference frame, with its centre of mass located at $\bm{x}_\textsc{a}$ and its spatial profile given by the real-valued distribution $F_\textsc{a}(\bm x)$. We denote by $t$ the proper time of the detector. The interaction Hamiltonian $\hat{H}_{\textsc{i},\textsc{a}}^{(1)}(t)$ between the detector and the field is (in the interaction picture)
\begin{equation}
\label{eq:H_I_A}
	\hat{H}_{\textsc{i},\textsc{a}}^{(1)}(t)
	=
	\lambda_\textsc{a} \chi_\textsc{a}(t) \hat{m}_\textsc{a}(t)
	\int \d[n]{\bm{x}} F_\textsc{a}(\bm{x}-\bm{x}_\textsc{a}) \hat{\phi}(\bm{x},t).
\end{equation}
Here, $\lambda_\textsc{a}$ is a coupling constant with units of $[\text{length}]^{(n-3)/2}$ in $n+1$ spacetime dimensions, and $\chi_\textsc{a}(t)$ is a dimensionless, time-dependent \textit{switching function}. $\hat{m}_\textsc{a}(t)$ is the monopole moment of the detector,
\begin{equation}
\label{eq:m_A}
	\hat{m}_\textsc{a}(t)
	=
	\ket{e_\textsc{a}}\bra{g_\textsc{a}} e^{\ii\Omega_\textsc{a} t}+
	\ket{g_\textsc{a}}\bra{e_\textsc{a}} e^{-\ii\Omega_\textsc{a} t}.
\end{equation}

The initial state $\hat{\rho}_\text{1,i}$ of the detector-field system evolves in time according to the unitary $\hat{U}_1$ generated by the interaction Hamiltonian~\eqref{eq:H_I_A}:
\begin{equation}
\label{eq:U_1}
	\hat{U}_1
	=
	\mathcal{T}\exp\left[{-\ii\int_{-\infty}^{\infty}\!\!\!\dif t\, \hat{H}_\textsc{i,a}^{(1)}(t)}\right],
\end{equation}
where $\mathcal{T}$ denotes the time-ordering operation. Hence the final state $\hat{\rho}_1$ of the system is 
\begin{equation}
\label{eq:rho_a_def}
	\hat{\rho}_1=
	\hat{U}_1 \hat{\rho}_\text{1,i} \hat{U}_1^\dagger,
\end{equation}
and the time evolved state of the detector, $\hat{\rho}_\textsc{a}$, is obtained by tracing out the field degrees of freedom:
\begin{equation}
\label{eq:rho_a_v1}
	\hat{\rho}_\textsc{a}
	=
	\Tr_{\hat{\phi}}(\hat{\rho}_1).
\end{equation}

\subsubsection{Two detectors}

We now consider two Unruh-DeWitt particle detectors, labelled $\nu\in\{\text{A}, \text{B}\}$, interacting with the field. We denote the initial time density matrix of the two detectors and the field by $\hat{\rho}_\text{2,i}$. Detector $\nu$ has ground state $\ket{g_\nu}$, excited state $\ket{e_\nu}$, and energy gap $\Omega_\nu$. We assume both detectors to be at rest in a common inertial reference frame, with the centre of mass of detector $\nu$ located at $\bm{x}_\nu$ and its spatial profile given by the real-valued distribution $F_\nu(\bm x)$. The coupling strength of detector $\nu$ is $\lambda_\nu$ and its time-dependent switching-function is $\chi_\nu(t)$. The interaction Hamiltonian $\hat{H}_\textsc{i,ab} (t)$ between the detectors and the field is, in the interaction picture,
\begin{equation}
\label{eq:H_I_AB}
	\hat{H}_\textsc{i,ab} (t)
	=
	\!\!\!\sum_{\nu\in\{\text{A,B}\}}
	\!\!\!
	\hat{H}_{\textsc{i},\nu}^{(2)}(t),
\end{equation}
with $\hat{H}_{\textsc{i},\nu}^{(2)}(t)$ given by
\begin{equation}
\label{eq:H_I_nu^2}
    \hat{H}_{\textsc{i},\nu}^{(2)}(t)
    =
    \lambda_\nu \chi_\nu(t) \hat{\mu}_\nu(t)
	\int \d[n]{\bm{x}} F_\nu(\bm{x}-\bm{x}_\nu) \hat{\phi}(\bm{x},t).
\end{equation}
Here, $\hat{\mu}_\nu(t)$ are operators acting on states in the two-detector Hilbert space, and are defined by
\begin{equation}
\label{eq:mu_nu}
	\hat{\mu}_\textsc{a}(t)
	=
	\hat{m}_\textsc{a}(t)\otimes\mathds{1}_\textsc{b},\quad 
	\hat{\mu}_\textsc{b}(t)
	=		\mathds{1}_\textsc{a}\otimes\hat{m}_\textsc{b}(t),
\end{equation}
where $\hat{m}_\nu$ is the monopole moment of detector $\nu$,
\begin{equation}
    \label{eq:m_nu}
	\hat{m}_\nu(t)=
	\ket{e_\nu}\bra{g_\nu} e^{\ii\Omega_\nu t}+
	\ket{g_\nu}\bra{e_\nu} e^{-\ii\Omega_\nu t}.
\end{equation}

As in the one-detector case, we write down the unitary $\hat{U}_2$ describing the evolution of the initial state $\hat{\rho}_\text{2,i}$ of the detectors-field system:
\begin{equation}
\label{eq:U2_def}
	\hat{U}_2
	=
	\mathcal{T}\exp\left[{-\ii\int_{-\infty}^{\infty}\!\!\!\dif t\, \hat{H}_\textsc{i,ab}(t)}\right].
\end{equation}
Hence the final state $\hat{\rho}_2$ of the detectors-field system is
\begin{equation}
	\hat{\rho}_2=
	\hat{U}_2 \hat{\rho}_\text{2,i} \hat{U}_2^\dagger,
\end{equation}
and the time evolved state of the two detectors, $\hat{\rho}_\textsc{ab}$, is obtained by tracing out the field degrees of freedom:
\begin{equation}
\label{eq:rho_ab_def}
	\hat{\rho}_\textsc{ab}
	=
	\Tr_{\hat{\phi}}(\hat{\rho}_2).
\end{equation}

\section{Single detector interacting with the field}
\label{sec:one_detector}

Let us now specify the initial conditions of the single detector and the field. We will assume that the field starts out in a general coherent state $\ket{\alpha(\bm k)}$, as defined in~\eqref{eq:alpha}, and that the detector is initialized to its free ground state. Thus the initial state $\hat{\rho}_\text{1,i}$ of the system is
\begin{equation}
\label{eq:rho_1i}
    \hat{\rho}_\text{1,i}
    =
    \ket{g_\textsc{a}}\bra{g_\textsc{a}}
    \otimes
    \ket{\alpha(\bm k)}\bra{\alpha(\bm k)}.
\end{equation}
From this, we wish to obtain an explicit expression for the time-evolved density matrix $\rhoa$~\eqref{eq:rho_a_v1} of a single detector following its interaction with the field. 

Problems of this type are often approached perturbatively, where calculations are only carried out to a certain order in the detector-field coupling strength $\lambda_\textsc{a}$, which is assumed to be small. For the case of detectors interacting with a coherent field state, this approach was taken in~\cite{Simidzija2017b}, to second order in $\lambda_\textsc{a}$.

In contrast, here we wish to obtain non-perturbative results. However, if our detector-field interaction Hamiltonian is non-zero at different times, then the presence of the time-ordering operation in the expression~\eqref{eq:U_1} for the time-evolution unitary $\hat{U}_1$ imposes mathematical challenges that are difficult to overcome. There is one simple way to bypass the time-ordering problem: we allow the detector to interact with the field at one instant in time (but with a finite amount of energy), using a delta-coupling~\cite{Hotta2008}. Although this may a priori seem an unphysical scenario, it is very physically relevant: it can be thought of as the limit of considering switching functions of finite area when the duration of the interaction is taken to be very short as compared to all other relevant scales in the problem.

More precisely, we first suppose that the detector interacts with the field according to a switching function $\chi_\textsc{a}^\sigma(t)$ that peaks at some time $t_\textsc{a}$, decays to zero away from $t_\textsc{a}$, and is symmetric about $t_\textsc{a}$. Suppose that $\chi_\textsc{a}^\sigma(t)$ has a characteristic width $\sigma$, i.e. the detector-field interaction is significant only for times $t$ such that $|t-t_\textsc{a}|<\sigma/2$. Finally, suppose that $\int_{-\infty}^\infty  \chi_\textsc{a}^\sigma(t) \dif t=\eta_\textsc{a}$, where $\eta_\textsc{a}$ has units of time. For example, $\chi_\textsc{a}^\sigma(t)$ could be a Gaussian, Lorentzian, or top-hat function. Then, taking the limit $\sigma\rightarrow 0$ while increasing the magnitude of $\chi_\textsc{a}^\sigma(t)$ near $t_\textsc{a}$ so that $\int_{-\infty}^\infty \chi_\textsc{a}^\sigma(t) \dif t$ remains constant, we obtain the detector switching function
\begin{equation}
\label{eq:chi_delta_a}
    \chi_\textsc{a}(t)
    =
    \eta_\textsc{a}\delta(t-t_\textsc{a}).
\end{equation}
This is the switching function that we will assume our detector obeys. We will soon see that this assumption allows us to obtain an explicit and non-perturbative expression for the evolved detector's density matrix $\rhoa$. It is important to note that we have also assumed that the regularization of this delta switching function is symmetric, i.e. that this delta switching is the symmetric limit of a symmetric switching function centered at $t_\textsc{a}$. Without this additional assumption our setup would be ambiguous, since different regularizations of a delta detector switching can yield different physical predictions~\cite{Pozas2017}. 

Assuming the detector switches on and off according to~\eqref{eq:chi_delta_a}, the expression for the unitary $\hat{U}_1$~\eqref{eq:U_1} simplifies to 
\begin{align}
\label{eq:U_1_v2}
    \hat{U}_1
    =
    \exp(\ma\otimes\Ya),
\end{align}
where from now on we take $\hat{m}_\textsc{a}$ to mean $\hat{m}_\textsc{a}(t_\textsc{a})$, and where $\Ya$ is defined as
\begin{equation}
\label{eq:Ya}
    \hat{Y}_\textsc{a}
    \coloneqq
    -\ii \lambda_\textsc{a}
    \eta_\textsc{a}
    \int\d[n]{\bm x}
    F_\textsc{a}(\bm{x}-\bm{x}_\textsc{a})
    \hat{\phi}(\bm{x},t_\textsc{a}).
\end{equation}
Note that, since $F_\textsc{a}(\bm x)$ is assumed to be real-valued and since the field $\hat{\phi}(\bm{x},t_\textsc{a})$ is a Hermitian operator, $\Ya$ is anti-Hermitian (i.e. $\Ya^\dagger=-\Ya$).

We would like to obtain a matrix representation of the time-evolved density matrix $\rhoa$. The obvious choice of basis for the detector's Hilbert space, in which we can express $\rhoa$, is \mbox{$\{\ket{g_\textsc{a}},\ket{e_\textsc{a}}\}$}. However, let us instead work in the orthonormal basis
\begin{equation}
\label{eq:B2}
    \{\ket{\tilde{g}_\textsc{a}},\ket{\tilde{e}_\textsc{a}}\},
\end{equation} 
where $\gak$ and $\eak$ are defined to be
\begin{align}
    \label{eq:gak}
    \ket{\tilde{g}_\textsc{a}}
    &\coloneqq
    \ket{g_\textsc{a}},
    \\
    \label{eq:eak}
    \ket{\tilde{e}_\textsc{a}}
    &\coloneqq
    e^{\ii\Omega_\textsc{a}t_\textsc{a}}
    \ket{e_\textsc{a}}.
\end{align}
Note that in the basis~\eqref{eq:B2}, the initial time density matrix of the detector is $\ket{\tilde{g}_\textsc{a}}\bra{\tilde{g}_\textsc{a}}$, while the detector's monopole moment is \mbox{$\ma=\eak\gab+\gak\eab$}. Since, in this basis, the representations of both of these matrices are independent of the detector's energy gap $\Omega_\textsc{a}$, this means that the representation of the evolved detector's density matrix $\rhoa$ will also be independent of $\Omega_\textsc{a}$ in this basis. Hence the eigenvalues of $\rhoa$, which are basis independent, are independent of the detector's energy gap.

Physically, we can think of this result as a consequence of the detector's delta coupling to the field: because the coupling is short and intense, the detector's free dynamics (governed by the size of the detector's energy gap) are overshadowed by the interaction-induced dynamics~\cite{Pozas2017}. In other words, the size of a detector's finite energy gap is irrelevant to the detector's dynamics if the detector interacts with the field in a pointlike manner in time.

To proceed, let us expand the exponential form of $\hat{U}_1$ in~\eqref{eq:U_1_v2} as a Taylor series and use the fact that $\ma^2=\mathds{1}_\textsc{a}$ is the identity operator on the detector's Hilbert space. We obtain
\begin{align}
    \hat{U}_1
    &=
    \sum_{n=0}^\infty
    \frac{1}{n!}
    (\ma\otimes\Ya)^n
    \notag\\
    &=
    \mathds{1}_\textsc{a}
    \otimes
    \sum_{n=0}^\infty
    \frac{1}{(2n)!}
    \Ya^{2n}
    +
    \ma
    \otimes
    \sum_{n=0}^\infty
    \frac{1}{(2n+1)!}
    \Ya^{2n+1}
    \notag\\
    &=
    \mathds{1}_\textsc{a}
    \otimes
    \cosh(\Ya)
    +
    \ma
    \otimes
    \sinh(\Ya).
\end{align}
Since $\ma$ is Hermitian and $\Ya$ is anti-Hermitian, we have that
\begin{equation}
    \hat{U}_1^\dagger
    =
    \mathds{1}_\textsc{a}
    \otimes
    \cosh(\Ya)
    -
    \ma
    \otimes
    \sinh(\Ya).
\end{equation}
Making use of these expressions for $\hat{U}_1$ and $\hat{U}_1^\dagger$, and noting that $\ma\gak=\eak$ and $\ma\eak=\gak$, the expression~\eqref{eq:rho_a_v1} for $\rhoa$ becomes
\begin{align}
\label{eq:rho_a_v2}
    \rhoa
    &=
    \tr_{\hat{\phi}}
    \big[
    \gak \gab \otimes
    \cosh(\Ya)\alphak\alphab\cosh(\Ya)
    \notag\\
    &\hspace{2.1em}-
    \gak \eab \otimes
    \cosh(\Ya)\alphak\alphab\sinh(\Ya)
    \notag\\
    &\hspace{2.1em}+
    \eak \gab \otimes
    \sinh(\Ya)\alphak\alphab\cosh(\Ya)
    \notag\\
    &\hspace{2.1em}-
    \eak \eab \otimes
    \sinh(\Ya)\alphak\alphab\sinh(\Ya)
    \big].
\end{align}
In the basis~\eqref{eq:B2} this becomes
\begin{equation}
\label{eq:rho_a_v3}
    \rhoa
    =
    \begin{pmatrix}
    f_1^{(++)} & f_1^{(-+)} \\
    f_1^{(+-)} & f_1^{(--)}
    \end{pmatrix},
\end{equation}
where we have defined $f_1^{(jk)}$ to be
\begin{align}
\label{eq:f}
    f_1^{(jk)}
    &\coloneqq
    \alphab
    \hat{X}_{(j)}^\dagger
    \hat{X}_{(k)}
    \alphak
    \notag\\
    &=
    \bra{0}\disp^\dagger
    \hat{X}_{(j)}^\dagger
    \hat{X}_{(k)}
    \disp\ket{0}.
\end{align}
The field operator $\hat{X}_{(j)}$ is defined as
\begin{align}
    \hat{X}_{(j)}
    &\coloneqq
    \frac{1}{2}
    \left(e^{\Ya}+je^{-\Ya}\right),
\end{align}
so that by $\hat{X}_{(j)}=\hat{X}_{(\pm)}$ we implicitly mean that $j=\pm 1$.

To obtain more useful expressions for the matrix elements $f_1^{(jk)}$, we need to commute the $\hat{X}_{(\pm)}$ operators past the displacement operator $\disp$. First, we consider the following commutator,
\begin{equation}
    \left[
    \Ya,
    \int\!\!\d[n]{\bm{k}}\!\!\left(\alpha(\bm{k})\ad{k}
	\!-\!
	\alpha^*(\bm{k})\a{k}\right)
	\right].
\end{equation}
Using expression~\eqref{eq:Ya} for $\Ya$ and making use of the canonical commutation relations, we find that this evaluates to $\ii\mathcal{C}_\textsc{a}\mathds{1}_{\hat{\phi}}$, where $\mathcal{C}_\textsc{a}$ is defined to be the real number
\begin{align}
\label{eq:c_a}
    \mathcal{C}_\textsc{a}
    \coloneqq
    -\lambda_\textsc{a}
    \eta_\textsc{a}\!
    \int\!\!
    \frac{\d[n]{\bm k}} {\sqrt{2|\bm{k}|}}
    \left(
    \tilde{F}_\textsc{a}(\bm k)
    \alpha(\bm k)
    e^{-\ii(|\bm k|t_\textsc{a} -\bm{k}\cdot \bm{x}_\textsc{a})}
    +\text{c.c.}
    \right),
\end{align}
and where $\tilde{F}_\textsc{a}(\bm k)$ is the Fourier transform of the detector's spatial profile $F_\textsc{a}(\bm x)$,
\begin{equation}
\label{eq:FT}
	\tilde{F}_\textsc{a}(\bm{k})
	\coloneqq
	\frac{1}{\sqrt{(2\pi)^n}}\int\d[n]{\bm{x}}
	F_\textsc{a}(\bm{x})e^{\ii\bm{k}\cdot\bm{x}}.
\end{equation}
Using the Baker-Campbell-Hausdorff formula it is now straightforward to show that
\begin{equation}
    \hat{X}_{(j)} \disp
    =
    \frac{1}{2}\disp
    \left(
    e^{\ii\ca}e^{\Ya} +j
    e^{-\ii\ca}e^{-\Ya}
    \right).
\end{equation}
This expression, along with its Hermitian adjoint, allows us to write $f_1^{(jk)}$ as
\begin{align}
\label{eq:f_v2}
    f_1^{(jk)}
    \!\!&=
    \!\frac{1}{4}
    \bra{0}\!\!
    \left(\!
    e^{-\ii\ca}e^{-\Ya}
    \!+\!j e^{\ii\ca}e^{\Ya}
    \right)\!\!
    \left(
    e^{\ii\ca}e^{\Ya}
    \!+\!k e^{-\ii\ca}e^{-\Ya}
    \right)\!\!
    \ket{0}
    \notag\\
    &=
    \frac{1}{4}
    \left(
    1+jk
    +je^{2\ii\ca}f_\textsc{a}
    +ke^{-2\ii\ca}f_\textsc{a}^*
    \right).
\end{align}
Here by, for example, $f_1^{(jk)}=f_1^{(+-)}$, we implicitly mean that $j=+1$ and $k=-1$. We have defined $f_\textsc{a}$ to be
\begin{equation}
\label{eq:fa}
    f_\textsc{a}
    \coloneqq
    \bra{0}e^{2\Ya}\ket{0},
\end{equation}
which is independent of the coherent amplitude of the field $\alpha(\bm k)$. Notice that if we define $\bnu$ to be
\begin{align}
\label{eq:bnu}
    \bnu
    &\coloneqq
    \frac{-2\ii\lambda_\nu\eta_\nu}{\sqrt{2(2\pi)^n|\bm{k}|}}
    \int \d[n]{\bm{x}}
    F_\nu(\bm{x}-\bm{x}_\nu)
    e^{\ii(|\bm{k}|t_\nu-\bm{k}\cdot\bm{x})}
    \notag\\
    &=
    \frac{-2\ii\lambda_\nu\eta_\nu}{\sqrt{2|\bm{k}|}}
    \tilde{F}_\nu(\bm{k})^*
    e^{\ii(|\bm{k}|t_\nu-\bm{k}\cdot\bm{x}_\nu)},
\end{align}
then $\fa$ can be rewritten as
\begin{equation}
    \fa
    =
    \langle 0 \ket{\ba},
\end{equation}
where $\ket{\ba}$ is a coherent state, as defined in equation~\eqref{eq:alpha}. Therefore $\fa$ is simply the inner product of a coherent state $\ket{\ba}$ and the field vacuum $\ket{0}$. In appendix~\ref{Appendix:coherent_inner_product} we show that this evaluates to
\begin{equation}
\label{eq:fa_final}
    f_\textsc{a}
    =
    \exp\left(
    -\frac{1}{2}
    \int\d[n]{\bm k}
    |\ba|^2
    \right),
\end{equation}
which is always real and strictly positive.

The evolved detector's density matrix $\rhoa$, the components of which are given in~\eqref{eq:f_v2}, can now be written as
\begin{equation}
\label{eq:rho_a_final}
	\rhoa
	=
	\frac{1}{2}
	\begin{pmatrix}
	1
	+\fa\cos
	\left(2\ca\right) &
	-\ii\fa\sin
	\left(2\ca\right) \\
	\ii\fa\sin
	\left(2\ca\right) &
	1
	-\fa\cos
	\left(2\ca\right)
	\end{pmatrix}.
\end{equation}

The density matrix $\rhoa$ in~\eqref{eq:rho_a_final} describes the state of a detector following its interaction with a coherent state of a scalar field, where we have assumed that the detector couples to the field according to a Dirac-delta type switching. We can compare this non-perturbative result to the one obtained in~\cite{Simidzija2017b} through a perturbative analysis. In~\cite{Simidzija2017b}, the same setup was considered (a detector interacting with a coherent scalar field state), but the detector was allowed to have an arbitrary switching function. In that study, a perturbative expression for $\rhoa$ to second order in the detector-field coupling strength $\lambda_\textsc{a}$ was obtained. As a consistency check, we show in appendix~\ref{Appendix:pert1} that these two calculations of $\rhoa$ agree in their common domain (delta switching, and to $\mathcal{O}(\lambda_\textsc{a}^2)$).

Next, let us compute the eigenvalues $E_{\textsc{a},i}$ of the density matrix $\rhoa$. We straightforwardly obtain the expressions
\begin{align}
\label{eq:E_A_1}
    E_{\textsc{a},1}
    &=
    \frac{1}{2}
    \left(1+\fa\right),
    \\
\label{eq:E_A_2}
    E_{\textsc{a},2}
    &=
    \frac{1}{2}
    \left(1-\fa\right).
\end{align}
Since the eigenvalues only depend on $f_\textsc{a}$, which is independent of the field's coherent amplitude $\alpha(\bm k)$, we arrive at the following result:

\vspace{2mm}
\textbf{Theorem 1:} 
\textit{Consider an UDW particle detector with arbitrary spatial smearing $F_\textsc{a}(\bm x)$, arbitrary field coupling strength $\lambda_\textsc{a}$, and delta switching function \mbox{$\chi_\textsc{a}(t)=\eta_\textsc{a}\delta(t-t_\textsc{a})$}. Then, the eigenvalues of the time evolved density matrix of the detector are the same whether the detector interacts with an arbitrary coherent state of a scalar field or with the vacuum.}

\textit{Therefore, any property of the time evolved state of the detector that depends solely on the spectrum of its density matrix is independent of whether the detector interacts with an arbitrary coherent state or with the scalar field vacuum.
}
\vspace{2mm}

This theorem is the non-perturbative analogue to Theorem 1 in~\cite{Simidzija2017b}, where it was shown that, to $\mathcal{O}(\lambda_\textsc{a}^2)$, a detector that couples to a coherent field state through an \textit{arbitrary} switching evolves into a state whose density matrix has a spectrum that is independent of $\alpha(\bm k)$.

To conclude this section, we give a physical consequence of our result. First, recall that the von Neumann entropy $S\left[\rhoa\right]$ of the state $\rhoa$ can be defined in terms of its eigenvalues as
\begin{equation}
    S\left[\rhoa\right]
    \coloneqq
    \sum_i 
    E_{\textsc{a},i}
    \ln
    \left(E_{\textsc{a},i}\right),
\end{equation}
and it is a monotonic measure of how entangled the detector and the field are (in particular, it is the field-detector entanglement entropy). Now suppose that we start with a particle detector in its ground state and the field in a coherent state, so that, initially, the detector and the field are unentangled. Then, if we allow the detector to interact with the coherent field state (e.g. we shine the detector with coherent light), the detector and the field will in general become entangled. What Theorem 1 tells us is that the field will become equally entangled with the detector regardless of which coherent state it is initially in. Furthermore, since the vacuum state is itself a coherent state (with vanishing coherent amplitude), a detector interacting with \textit{any} coherent field state will become just as entangled with the field as if it interacted with the vacuum.

\section{Two detectors interacting with the field}
\label{sec:two_detectors}

As in the single detector case, we assume that the two detectors are initially in their ground states, and that the field is in an arbitrary coherent state. Hence the initial state $\hat{\rho}_\text{2,i}$ of the system is
\begin{equation}
\label{eq:rho_2i}
    \hat{\rho}_\text{2,i}
    =
    \ket{g_\textsc{a}}\bra{g_\textsc{a}}
    \otimes
    \ket{g_\textsc{b}}\bra{g_\textsc{b}}
    \otimes
    \ket{\alpha(\bm k)}\bra{\alpha(\bm k)}.
\end{equation}
From this, we wish to obtain an expression for the time-evolved, two detector density matrix $\rhoab$~\eqref{eq:rho_ab_def}. To obtain a non-perturbative expression for $\rhoab$, we follow the same procedure as in the single detector case. First, we suppose that detector $\nu$ interact with the field according to a (symmetric) switching function centered at time $t_\nu$, and then we take the limit so that the interaction is infinitesimally short. In other words, we set the detector switching function to be
\begin{equation}
\label{eq:chi_delta_nu}
    \chi_\nu(t)
    =
    \eta_\nu\delta(t-t_\nu),
\end{equation}
and we also make the assumption that this expression is the symmetrically taken limit of a symmetric regularization function (again, as in \cite{Pozas2017}). Here, $\eta_\nu$ are constants with dimensions of time, so that $\chi_\nu(t)$ are dimensionless functions. With the above switching functions, the unitary $\hat{U}_2$ (defined in~\eqref{eq:U2_def}) governing the time evolution of the detectors-field system takes the form
\begin{equation}
    \hat{U}_2
    =
    \mathcal{T}
    \exp
    \left(
    -\ii 
    \left[
    \hat{H}_\textsc{i,a}^{(2)}(t_\textsc{a})
    +\hat{H}_\textsc{i,b}^{(2)}(t_\textsc{b})
    \right]
    \right).
\end{equation}
Without loss of generality, let us assume that detector A interacts with the field no later than detector B. That is we assume $t_\textsc{a}\le t_\textsc{b}$. Under this condition, the expression for $\hat{U}_2$ simplifies to (see appendix~\ref{Appendix:U2} for details)
\begin{align}
\label{eq:U2}
    \hat{U}_2
    &=
    \exp
    \left(-\ii 
    \hat{H}_\textsc{i,b}^{(2)}(t_\textsc{b})
    \right)
    \exp
    \left(-\ii 
    \hat{H}_\textsc{i,a}^{(2)}(t_\textsc{a})
    \right).
\end{align}
We can alternatively write this expression as
\begin{equation}
    \hat{U}_2
    =
    \exp
    \left(
    \mub
    \otimes
    \Yb
    \right)
    \exp
    \left(
    \mua
    \otimes
    \Ya
    \right).
\end{equation}
Here by $\hat{\mu}_\nu$ we implicitly mean $\hat{\mu}_\nu(t_\nu)$---as in eq.~\eqref{eq:mu_nu}---which is an operator on the two-detector Hilbert space. Meanwhile $\Ya$, defined in~\eqref{eq:Ya}, and $\Yb$, defined analogously as
\begin{equation}
\label{eq:Yb}
    \hat{Y}_\textsc{b}
    \coloneqq
    -\ii \lambda_\textsc{b}
    \eta_\textsc{b}
    \int\d[n]{\bm x}
    F_\textsc{b}(\bm{x}-\bm{x}_\textsc{b})
    \hat{\phi}(\bm{x},t_\textsc{b}),
\end{equation}
are operators on the field's Hilbert space. Using the fact that $\hat{m}_\nu^2=\mathds{1}_\nu$ is the identity operator on the Hilbert space of detector $\nu$ allows us to write $\hat{U}_2$ as
\begin{align}
\label{eq:U2_v2}
    \hat{U}_2
    &=
    \left(
    \mathds{1}_\textsc{a}
    \otimes
    \mathds{1}_\textsc{b}
    \otimes
    \cosh(\Yb)
    +
    \mathds{1}_\textsc{a}
    \otimes
    \mb
    \otimes
    \sinh(\Yb)
    \right)
    \notag\\
    &\hspace{1.2em}\times
    \left(
    \mathds{1}_\textsc{a}
    \otimes
    \mathds{1}_\textsc{b}
    \otimes
    \cosh(\Ya)
    +
    \ma
    \otimes
    \mathds{1}_\textsc{b}
    \otimes
    \sinh(\Ya)
    \right)
    \notag\\
    &=
    \mathds{1}_\textsc{a}
    \otimes
    \mathds{1}_\textsc{b}
    \otimes
    \hat{X}_{(++)}
    +
    \ma
    \otimes
    \mathds{1}_\textsc{b}
    \otimes
    \hat{X}_{(+-)}
    \notag\\
    &\hspace{1.2em}
    +
    \mathds{1}_\textsc{a}
    \otimes
    \mb
    \otimes
    \hat{X}_{(-+)}
    +
    \ma
    \otimes
    \mb
    \otimes
    \hat{X}_{(--)},
\end{align}
where the field operators $\hat{X}_{(jk)}$ are defined to be
\begin{equation}
\label{eq:X_jk}
    \hat{X}_{(jk)}
    \coloneqq
    \frac{1}{4}
    \left(
    e^{\Yb}+j e^{-\Yb}
    \right)
    \left(
    e^{\Ya}+k e^{-\Ya}
    \right),
\end{equation}
and where by, for example, $\hat{X}_{(jk)}=\hat{X}_{(-+)}$ we implicitly mean that $j=-1$ and $k=+1$.

Following the same logic as in the one-detector case, let us define the following convenient orthonormal basis for the two detector Hilbert space,
\begin{equation}
\label{eq:B2_2}
    \{
    \gak\otimes\gbk,
    \gak\otimes\ebk,
    \eak\otimes\gbk,
    \eak\otimes\ebk
    \},
\end{equation}
where $\ket{\tilde{g}_\nu}$ and $\ket{\tilde{e}_\nu}$ are defined to be
\begin{align}
    \label{eq:gnuk}
    \ket{\tilde{g}_\nu}
    &\coloneqq
    \ket{g_\nu},
    \\
    \label{eq:enuk}
    \ket{\tilde{e}_\nu}
    &\coloneqq
    e^{\ii\Omega_\nu t_\nu}
    \ket{e_\nu}.
\end{align}
With these definitions we can write the initial state of the detector-field system as
\begin{equation}
\label{eq:rho_2i_tilde}
    \hat{\rho}_\text{2,i}
    =
    \gak\gab
    \otimes
    \gbk\gbb
    \otimes
    \ket{\alpha(\bm k)}\bra{\alpha(\bm k)}.
\end{equation}
Hence, in the basis~\eqref{eq:B2_2}, the matrix elements of \mbox{$\mathds{1}_\textsc{a}\otimes\mathds{1}_\textsc{b}$}, $\ma\otimes\mathds{1}_\textsc{b}$, $\mathds{1}_\textsc{a}\otimes\mb$ and $\ma\otimes\mb$, as well as the matrix elements of the initial two-detector density matrix, are all independent of the detectors' energy gaps $\Omega_\textsc{a}$ and $\Omega_\textsc{b}$. Therefore, the matrix elements of the time-evolved two detector density matrix $\rhoab$ are also independent of the detectors' energy gaps, as is any basis independent property of $\rhoab$ (such as its spectrum). We will work in the basis~\eqref{eq:B2_2} so that we can explicitly see this independence (and so that we simplify the expressions for the matrix elements of $\rhoab$). We find that $\rhoab$ in~\eqref{eq:rho_ab_def} takes the form
\begin{equation}
\label{eq:rhoab}
    \rhoab
    =
    \begin{pmatrix}
    f_2^{(++++)} & f_2^{(-+++)} & f_2^{(+-++)} & f_2^{(--++)}
    \\
    f_2^{(++-+)} & f_2^{(-+-+)} & f_2^{(+--+)} & f_2^{(---+)}
    \\
    f_2^{(+++-)} & f_2^{(-++-)} & f_2^{(+-+-)} & f_2^{(--+-)}
    \\
    f_2^{(++--)} & f_2^{(-+--)} & f_2^{(+---)} & f_2^{(----)}
    \end{pmatrix},
\end{equation}
where we have defined $f_2^{(jklm)}$ to be
\begin{align}
\label{eq:f2_def}
    f_2^{(jklm)}
    &\coloneqq
    \alphab
    \hat{X}_{(jk)}^\dagger
    \hat{X}_{(lm)}
    \alphak
    \notag\\
    &=
    \bra{0}\disp^\dagger
    \hat{X}_{(jk)}^\dagger
    \hat{X}_{(lm)}
    \disp\ket{0}.
\end{align}
To further simplify this expression, we first recall the definition of $\ca$ in~\eqref{eq:c_a}, and define $\cb$ analogously,
\begin{equation}
\label{eq:c_b}
    \cb
    \coloneqq
    -\lambda_\textsc{b}
    \eta_\textsc{b}\!
    \int\!\!
    \frac{\d[n]{\bm k}} {\sqrt{2|\bm{k}|}}
    \left(
    \tilde{F}_\textsc{b}(\bm k)
    \alpha(\bm k)
    e^{-\ii(|\bm k|t_\textsc{b} -\bm{k}\cdot \bm{x}_\textsc{b})}
    +\text{c.c.}
    \right),
\end{equation}
so that we have
\begin{align}
    \left[
    \Yb,
    \int\!\!\d[n]{\bm{k}}\!\!\left(\alpha(\bm{k})\ad{k}
	\!-\!
	\alpha^*(\bm{k})\a{k}\right)
	\right]=\ii\cb \mathds{1}_{\hat{\phi}}.
\end{align}
After some calculations (see appendix~\ref{Appendix:f2} for details), we obtain the following expression for the matrix components $f_2^{(jklm)}$ of $\rhoab$:
\begin{align}
\label{eq:f2}
    f_2^{(jklm)}
    &=
    \frac{1}{16}
    \big[
    1+jl+km+jklm
    \notag\\
    &+l\fb^*(kme^{-2\ii\theta}+e^{2\ii\theta}) e^{-2\ii\cb}
    +m\fa^*(1+jl) e^{-2\ii\ca}
    \notag\\ 
    &+j\fb(e^{-2\ii\theta}+kme^{2\ii\theta}) e^{2\ii\cb}
    +k\fa(1+jl) e^{2\ii\ca}
    \notag\\
    &+jk\fp e^{2\ii\theta} e^{2\ii\ca}e^{2\ii\cb}
    +ml\fp^*e^{-2\ii\theta} e^{-2\ii\ca}e^{-2\ii\cb}
    \notag\\
    &+jm\fm e^{-2\ii\theta} e^{-2\ii\ca}e^{2\ii\cb}
    +kl\fm^* e^{2\ii\theta} e^{2\ii\ca}e^{-2\ii\cb}
    \big],
\end{align}
where by, for example, $f_2^{(jklm)}=f_2^{(++--)}$ we implicitly mean that $j=k=+1$ and $l=m=-1$. Here, $\theta$ is defined to be the real number
\begin{align}
\label{eq:theta}
    \theta
    \coloneqq\,
    & \ii\lambda_\textsc{a}
    \lambda_\textsc{b}
    \eta_\textsc{a}
    \eta_\textsc{b}
    \int\frac{\d[n]{\bm k}}{2|\bm{k}|}
    \notag\\
    &\times
    \left(
    \tilde{F}_\textsc{a}(\bm k)
    \tilde{F}_\textsc{b}^*(\bm k)
    e^{-\ii|\bm{k}|(t_\textsc{a}-t_\textsc{b})}
    e^{\ii\bm{k}\cdot (\bm{x}_\textsc{a}-\bm{x}_\textsc{b})}
    -\text{c.c.}
    \right),
\end{align}
so that $[\Ya,\Yb]=\ii\theta\mathds{1}_{\hat{\phi}}$. Meanwhile, $\fa$ was defined in~\eqref{eq:fa}, while $\fb$, $\fp$, and $\fm$ are defined to be
\begin{align}
\label{eq:fb}
    \fb
    &\coloneqq
    \bra{0}e^{2\Yb}\ket{0},
    \\
\label{eq:fp}
    \fp
    &\coloneqq
    \bra{0}e^{2\Yb}e^{2\Ya}\ket{0},
    \\
\label{eq:fm}
    \fm
    &\coloneqq
    \bra{0}e^{2\Yb}e^{-2\Ya}\ket{0}.
\end{align}
Notice that, as we already saw to be the case for $\fa$, the above expressions are also inner products between coherent states. Namely:
\begin{align}
    f_\textsc{b}
    &=
    \bra{0} \bb\rangle,
    \\
    f_\text{p}
    &=
    \bra{-\bb} \ba \rangle,
    \\
    f_\text{m}
    &=
    \bra{-\bb} -\ba
    \rangle,
\end{align}
where $\bnu$ is defined in~\eqref{eq:bnu}. In appendix~\ref{Appendix:coherent_inner_product} we show that the inner product between two arbitrary coherent states, $\ket{\ba}$ and $\ket{\bb}$, is
\begin{align}
    &\langle\bb\ket{\ba}
    \notag\\
    =&
    \exp \!
    \left[
    -\frac{1}{2}
    \int\!\!\d[n]{\bm k}\!
    \left(
    |\ba|^2+
    |\bb|^2-
    2\ba\bb^*
    \right)
    \right]\!.
\end{align}
This allows us to write $\fb$, $\fp$ and $\fm$ in the more convenient forms,
\begin{align}
\label{eq:fb_final}
    \fb
    &=
    \exp\left(
    -\frac{1}{2}
    \int\!\!\d[n]{\bm k}
    |\bb|^2
    \right),
    \\
\label{eq:fp_final}
    f_\text{p}
    &=
    \exp\!\Bigg[\!\frac{-1}{2}
    \!\!\!\int\!\!\d[n]{\bm k}
    \!\Big(\!
    |\ba|^2
    +|\bb|^2
    +2\ba\bb^*
    \!\Big)\!\Bigg]\!,
    \\
\label{eq:fm_final}
    f_\text{m}
    &=
    \exp\!\Bigg[\!\frac{-1}{2}
    \!\!\!\int\!\!\d[n]{\bm k}
    \!\Big(\!
    |\ba|^2
    +|\bb|^2
    -2\ba\bb^*
    \!\Big)\!\Bigg]\!.
\end{align}
Notice that $\fb$ (like $\fa$) is real, while $\fp$ and $\fm$ are, in general, complex.

Substituting~\eqref{eq:f2} into~\eqref{eq:rhoab}, we obtain an explicit matrix representation for the density matrix $\rhoab$, which describes the state of the two detectors following their interactions with a coherent state of the scalar field. In~\cite{Simidzija2017b}, a perturbative expression for $\rhoab$ was obtained (to second order in the detectors' coupling strengths) for detectors that couple to the field through an arbitrary switching function. As a consistency check, in appendix~\ref{Appendix:pert2} we verify that these two expressions of $\rhoab$ are consistent with one another when we particularize our exact result to the perturbative regime.

It is difficult to learn anything meaningful from the current expression for $\rhoab$ in~\eqref{eq:rhoab}. Notice however, that the matrix elements $f_2^{(jklm)}$~\eqref{eq:f2} of $\rhoab$ only depend on the field's coherent amplitude $\alpha(\bm k)$ through $\ca$ and $\cb$. This motivates factoring $\rhoab$ as
\begin{equation}
\label{eq:rho_ab_fact}
    \rhoab
    =
    \hat{W}^\dagger \hat{Q} \hat{W},
\end{equation}
where $\hat{W}$ and $\hat{Q}$ are defined to be
\begin{align}
\label{eq:W}
    \hat{W}\!
    &\coloneqq
    \frac{1}{2}\!
    \begin{pmatrix}
    e^{-\ii\ca}e^{-\ii\cb} &
    e^{-\ii\ca}e^{-\ii\cb} &
    e^{-\ii\ca}e^{-\ii\cb} &
    e^{-\ii\ca}e^{-\ii\cb} 
    \\
    e^{-\ii\ca}e^{\ii\cb} &
    -e^{-\ii\ca}e^{\ii\cb} &
    e^{-\ii\ca}e^{\ii\cb} &
    -e^{-\ii\ca}e^{\ii\cb} 
    \\
    e^{\ii\ca}e^{-\ii\cb} &
    e^{\ii\ca}e^{-\ii\cb} &
    -e^{\ii\ca}e^{-\ii\cb} &
    -e^{\ii\ca}e^{-\ii\cb} 
    \\
    e^{\ii\ca}e^{\ii\cb} &
    -e^{\ii\ca}e^{\ii\cb} &
    -e^{\ii\ca}e^{\ii\cb} &
    e^{\ii\ca}e^{\ii\cb}
    \end{pmatrix}\!\!,
    \\
\label{eq:Q}
    \hat{Q}
    &\coloneqq
    \frac{1}{4}\!
    \begin{pmatrix}
    1 & e^{-2\ii\theta} \fb & \fa & e^{2\ii\theta} \fp
    \\
    e^{2\ii\theta} \fb & 1 & e^{2\ii\theta} \fm^* & \fa
    \\
    \fa & e^{-2\ii\theta} \fm & 1 & e^{2\ii\theta} \fb
    \\
    e^{-2\ii\theta} \fp^* & \fa & e^{-2\ii\theta} \fb & 1
    \end{pmatrix}.
\end{align}
Notice that $\hat{W}$ is unitary, and that $\hat{Q}$ is independent of $\ca$ and $\cb$. Hence $\rhoab$ is unitarily equivalent to a matrix $\hat{Q}$, which (in a certain basis) has components that are independent of the field's coherent amplitude $\alpha(\bm k)$.

Now, let us consider the partial transpose with respect to system A of the matrix $\rhoab$ which we denote $\rhoabpt$. In the basis~\eqref{eq:B2_2} this takes the form
\begin{equation}
\label{eq:rhoabpt}
    \rhoabpt
    =
    \begin{pmatrix}
    f_2^{(++++)} & f_2^{(-+++)} & f_2^{(+++-)} & f_2^{(-++-)}
    \\
    f_2^{(++-+)} & f_2^{(-+-+)} & f_2^{(++--)} & f_2^{(-+--)}
    \\
    f_2^{(+-++)} & f_2^{(--++)} & f_2^{(+-+-)} & f_2^{(--+-)}
    \\
    f_2^{(+--+)} & f_2^{(---+)} & f_2^{(+---)} & f_2^{(----)}
    \end{pmatrix},
\end{equation}
and can be factored as
\begin{equation}
\label{eq:rho_abpt_fact}
    \rhoabpt
    =
    \hat{V}^\dagger
    \hat{Q}^{{\text{\textbf{t}}}_\textsc{a}} 
    \hat{V}.
\end{equation}
Here, $\hat{Q}^{{\text{\textbf{t}}}_\textsc{a}}$ is the partial transpose of $\hat{Q}$, 
\begin{equation}
\label{eq:Qpt}
    \hat{Q}^{{\text{\textbf{t}}}_\textsc{a}}
    =
    \frac{1}{4}
    \begin{pmatrix}
    1 & e^{-2\ii\theta} \fb & \fa & e^{-2\ii\theta} \fm
    \\
    e^{2\ii\theta} \fb & 1 & e^{-2\ii\theta} \fp^* & \fa
    \\
    \fa & e^{2\ii\theta} \fp & 1 & e^{2\ii\theta} \fb
    \\
    e^{2\ii\theta} \fm^* & \fa & e^{-2\ii\theta} \fb & 1
    \end{pmatrix},
\end{equation}
and $\hat{V}$ is defined to be
\begin{align}
\label{eq:V}
    \hat{V}\!
    &\coloneqq
    \!\frac{1}{2}\!\!
    \begin{pmatrix}
    e^{\ii\ca}e^{-\ii\cb} &
    e^{\ii\ca}e^{-\ii\cb} &
    e^{\ii\ca}e^{-\ii\cb} &
    e^{\ii\ca}e^{-\ii\cb} 
    \\
    e^{\ii\ca}e^{\ii\cb} &
    -e^{\ii\ca}e^{\ii\cb} &
    e^{\ii\ca}e^{\ii\cb} &
    -e^{\ii\ca}e^{\ii\cb} 
    \\
    e^{-\ii\ca}e^{-\ii\cb} &
    e^{-\ii\ca}e^{-\ii\cb} &
    -e^{\!-\ii\ca}e^{\!-\ii\cb} &
    -e^{\!-\ii\ca}e^{\!-\ii\cb} 
    \\
    e^{-\ii\ca}e^{\ii\cb} &
    -e^{-\ii\ca}e^{\ii\cb} &
    -e^{-\ii\ca}e^{\ii\cb} &
    e^{-\ii\ca}e^{\ii\cb}
    \end{pmatrix}\!\!.
\end{align}
Notice that $\hat{V}$ is unitary, and that $\hat{Q}^{{\text{\textbf{t}}}_\textsc{a}}$ is independent of $\ca$ and $\cb$. Hence $\rhoab$ is unitarily equivalent to a matrix $\hat{Q}^{{\text{\textbf{t}}}_\textsc{a}}$, which (in a certain basis) has components that are independent of the field's coherent amplitude $\alpha(\bm k)$. Since unitarily equivalent matrices have the same eigenvalues, we come to the following conclusion:

\vspace{2mm}
\textbf{Theorem 2:}
\textit{
Consider two UDW particle detectors $\mathrm{(A}$ and $\mathrm{B)}$ with arbitrary spatial smearings $F_\nu(\bm x)$, arbitrary field coupling strengths $\lambda_\nu$, and delta switching functions \mbox{$\chi_\nu(t)=\eta_\nu\delta(t-t_\nu)$}, where $\nu\in\{\mathrm{A},\mathrm{B} \}$. Then, the eigenvalues of the time evolved density matrix of the two detectors are the same whether the detectors interact with an arbitrary coherent state of a scalar field or with the vacuum. Additionally, the eigenvalues of the partially-transposed density matrix of the two detectors are also the same whether the detectors interact with an arbitrary coherent state or with the vacuum.}

\textit{Therefore, any property of the time evolved state of the detectors that depends solely on the spectrum of their density matrix, or of its partial transpose, is independent of whether the detectors interact with an arbitrary coherent state or with the scalar field vacuum.}
\vspace{2mm}

This result is the non-perturbative analogue to that obtained in~\cite{Simidzija2017b},
where it was shown that, for detectors obeying \textit{arbitrary} switching functions, the spectra of $\rhoab$ and $\rhoabpt$ are independent of the field's coherent amplitude, to second order in the detector-field coupling strengths. 

We end this section by discussing one physical consequence of Theorem 2. First, we note that the \textit{negativity} $\mathcal{N}\left[\rhoab\right]$ of a state $\rhoab$ is defined as~\cite{Vidal2002}
\begin{equation}
\label{eq:neg}
    \mathcal{N}\left[\rhoab\right]
    \coloneqq
    \sum_i
    \max
    \left(0,-E_{\textsc{ab},i}^{{\text{\textbf{t}}}_\textsc{a}}
    \right),
\end{equation}
where the $E_{\textsc{ab},i}^{{\text{\textbf{t}}}_\textsc{a}}$ are the eigenvalues of the partially transposed matrix $\rhoabpt$. It has been shown that the negativity of a two-qubit system is an entanglement monotone that vanishes if and only if the two-qubit state is separable~\cite{Peres1996,Horodecki1996}. Hence the negativity is often used as a measure of entanglement in harvesting scenarios.

In our setup, we considered two comoving particle detectors that are initially in their ground states, and are therefore initially separable. Then, we supposed that detector $\nu$ interacts with a coherent field state (e.g. we shine it with coherent light) at time $t_\nu$. Following the interactions the detectors evolve into some final state. What Theorem 2 tells us is that the amount of entanglement in the evolved two-detector state is independent of the particular modes present in the coherent light beam that the detectors interacted with. Furthermore, since the vacuum state of the field is itself a coherent state (with vanishing coherent amplitude), this implies that the detectors will harvest the same amount of entanglement from \textit{any} coherent state as from the field vacuum.

\section{Entanglement harvesting}
\label{sec:ent_harv}

We saw in the previous section that a pair of detectors---initially in their ground states and coupling to the field through delta switching functions---can harvest the same amount of entanglement from any coherent field state as they can from the vacuum. However, we did not determine what this amount of entanglement is. That is the focus of this section.

Through a perturbative analysis, it was shown in~\cite{Pozas2017} that to second order in the detector-field coupling strengths, detectors that couple to the field according to delta switching functions \textit{cannot} harvest entanglement from the field vacuum. Furthermore, combining this result with the results in~\cite{Simidzija2017b}, this is therefore also true if the field is in any coherent state. We would like to see whether these results also hold in the non-perturbative regime.

First, let us express the matrices $\hat{Q}$~\eqref{eq:Q} and $\hat{Q}^{{\text{\textbf{t}}}_\textsc{a}}$~\eqref{eq:Qpt} in a different form. Notice that $\hat{Q}$ and $\hat{Q}^{{\text{\textbf{t}}}_\textsc{a}}$ are written in terms of five functions ($\fa$, $\fb$, $f_\text{p}$, $f_\text{m}$, and $\theta$), and that in equations \eqref{eq:fa_final}, \eqref{eq:fb_final}, \eqref{eq:fp_final}, and \eqref{eq:fm_final}, we have written the first four of these functions in terms of $\ba$ and $\bb$, which are defined in~\eqref{eq:bnu}. We can also express $\theta$, defined in~\eqref{eq:theta}, in terms of $\ba$ and $\bb$. Namely,
\begin{equation}
\label{eq:theta_final}
    \theta
    =
    \frac{\ii}{4}\int\d[n]{\bm k}
    \left(
    \ba^*\bb
    -\text{c.c}
    \right).
\end{equation}
We will also find it useful to define the following real number, which we denote $\omega$:
\begin{equation}
\label{eq:omega}
    \omega 
    \coloneqq
    -\frac{1}{2}\int\d[n]{\bm k}
    \left(
    \ba^*\bb
    +\text{c.c}
    \right).
\end{equation}
Hence we see that $f_\text{p}$ and $f_\text{m}$ can be expressed as
\begin{align}
    f_\text{p}
    &=f_\textsc{a}f_\textsc{b}
    e^{\omega-2\ii\theta},
    \\
    f_\text{m}
    &=f_\textsc{a}f_\textsc{b}
    e^{-\omega+2\ii\theta}.
\end{align}
Writing $f_\text{p}$ and $f_\text{m}$ in this form, and noting that $f_\textsc{a}$ and $f_\textsc{b}$ are real, allows us to write the matrices $\hat{Q}$ and $\hat{Q}^{{\text{\textbf{t}}}_\textsc{a}}$ as
\begin{align}
\label{eq:Q2}
    \hat{Q}
    &=
    \frac{1}{4}
    \begin{pmatrix}
    1 & e^{-2\ii\theta} \fb & \fa & \fa\fb e^\omega
    \\
    e^{2\ii\theta} \fb & 1 & \fa\fb e^{-\omega} & \fa
    \\
    \fa & \fa\fb e^{-\omega} & 1 & e^{2\ii\theta} \fb
    \\
    \fa\fb e^\omega & \fa & e^{-2\ii\theta} \fb & 1
    \end{pmatrix},
    \\
\label{eq:Qpt2}
    \hat{Q}^{{\text{\textbf{t}}}_\textsc{a}}
    &=
    \frac{1}{4}
    \begin{pmatrix}
    1 & e^{-2\ii\theta} \fb & \fa & \fa\fb e^{-\omega}
    \\
    e^{2\ii\theta} \fb & 1 & \fa\fb e^{\omega} & \fa
    \\
    \fa & \fa\fb e^{\omega} & 1 & e^{2\ii\theta} \fb
    \\
    \fa\fb e^{-\omega} & \fa & e^{-2\ii\theta} \fb & 1
    \end{pmatrix}.
\end{align}
Notice that $\hat{Q}$ and $\hat{Q}^{{\text{\textbf{t}}}_\textsc{a}}$ are equivalent up to a change in the sign of $\omega$.

Now recall from the previous section that we can quantify the amount of entanglement in the two-detector state $\rhoab$ by its negativity $\mathcal{N}\left[\rhoab\right]$, defined in equation~\eqref{eq:neg}. To determine the negativity, we first need to determine the eigenvalues $E_{\textsc{ab},i}^{{\text{\textbf{t}}}_\textsc{a}}$ of the partially transposed density matrix $\rhoabpt$. We showed in the previous section that these are the same as the eigenvalues of $\hat{Q}^{{\text{\textbf{t}}}_\textsc{a}}$, which is defined in~\eqref{eq:Qpt2} and is significantly easier to handle than $\rhoabpt$. We can straightforwardly calculate the spectrum of $\hat{Q}^{{\text{\textbf{t}}}_\textsc{a}}$, obtaining
\begin{align}
\label{eq:e1}
    E_{\textsc{ab},1}^{{\text{\textbf{t}}}_\textsc{a}}&=
    \frac{1}{8}
    \Big[
    2-\left(e^\omega +e^{-\omega}\right)\fa\fb
    \\
    &\hspace{2.2em}+\sqrt{
    4\left|\fa e^{\ii\theta}-\fb e^{-\ii\theta}\right|^2 +
    \left(e^\omega -e^{-\omega}\right)^2 \fa^2\fb^2
    }
    \Big],
    \notag\\
\label{eq:e2}
    E_{\textsc{ab},2}^{{\text{\textbf{t}}}_\textsc{a}}&=
    \frac{1}{8}
    \Big[
    2-\left(e^\omega +e^{-\omega}\right)\fa\fb
    \\
    &\hspace{2.2em}-\sqrt{
    4\left|\fa e^{\ii\theta}-\fb e^{-\ii\theta}\right|^2 +
    \left(e^\omega -e^{-\omega}\right)^2 \fa^2\fb^2
    }
    \Big],
    \notag\\
\label{eq:e3}
    E_{\textsc{ab},3}^{{\text{\textbf{t}}}_\textsc{a}}&=
    \frac{1}{8}
    \Big[
    2+\left(e^\omega +e^{-\omega}\right)\fa\fb
    \\
    &\hspace{2.2em}+\sqrt{
    4\left|\fa e^{\ii\theta}+\fb e^{-\ii\theta}\right|^2 +
    \left(e^\omega -e^{-\omega}\right)^2 \fa^2\fb^2
    }
    \Big],
    \notag\\
\label{eq:e4}
    E_{\textsc{ab},4}^{{\text{\textbf{t}}}_\textsc{a}}&=
    \frac{1}{8}
    \Big[
    2+\left(e^\omega +e^{-\omega}\right)\fa\fb
    \\
    &\hspace{2.2em}-\sqrt{
    4\left|\fa e^{\ii\theta}+\fb e^{-\ii\theta}\right|^2 +
    \left(e^\omega -e^{-\omega}\right)^2 \fa^2\fb^2
    }
    \Big].
    \notag
\end{align}

Here we notice something remarkable: the eigenvalues $E_{\textsc{ab},i}^{{\text{\textbf{t}}}_\textsc{a}}$ of $\rhoabpt$ and $\hat{Q}^{{\text{\textbf{t}}}_\textsc{a}}$ are invariant with respect to the transformation \mbox{$\omega\rightarrow -\omega$}. However, we already saw that this transformation changes $\hat{Q}^{{\text{\textbf{t}}}_\textsc{a}}$ into $\hat{Q}$.

Hence, the $E_{\textsc{ab},i}^{{\text{\textbf{t}}}_\textsc{a}}$ are also the eigenvalues of $\hat{Q}$, and therefore of $\rhoab$. Furthermore, since $\rhoab$ is a density matrix, it must have non-negative eigenvalues. Thus we conclude that the eigenvalues of the partially transposed density matrix $\rhoabpt$ are non-negative as well. In any case, and for completeness, in appendix~\ref{Appendix:eigenvalues} we provide an explicit proof that all of the eigenvalues $E_{\textsc{ab},i}^{{\text{\textbf{t}}}_\textsc{a}}$ of $\rhoabpt$ are non-negative.

We have therefore shown that the eigenvalues of the partially transposed two-detector density matrix $\rhoabpt$ are always non-negative, and hence that the negativity $\mathcal{N}\left[\rhoab\right]=0$. This results in the following conclusion:

\vspace{2mm}
\textbf{Theorem 3:} \textit{Consider two UDW particle detectors $\mathrm{(A}$ and $\mathrm{B)}$ with arbitrary spatial smearings $F_\nu(\bm x)$, arbitrary field coupling strengths $\lambda_\nu$, and delta switching functions \mbox{$\chi_\nu(t)=\eta_\nu\delta(t-t_\nu)$}, where $\nu\in\{\mathrm{A},\mathrm{B} \}$. Then, if we allow the detectors to interact with a coherent state of the scalar field (which in particular includes the case of the vacuum state), following the interaction they will remain in a separable state: i.e. the detector pair cannot harvest entanglement from the field.}
\vspace{2mm}

There are several possible ways to gain some physical insight on this result based on previous literature.

First, in~\cite{Reznik2005} (and further explored in \cite{Pozas2015}), it was shown that the amount of entanglement that a detector pair can harvest from the vacuum state of a scalar field (as measured by the negativity) is a direct competition between non-local terms (that increase the negativity), and local terms (that decrease it). This may agree with one's intuition: for a two-detector state to be entangled, the non-local correlations between the detectors must overcome the noises affecting each detector locally. It was discussed in \cite{Pozas2015} that the more sudden a switching is, the more the local noise can overwhelm the correlation terms. If we extend this intuition to entanglement harvesting from arbitrary coherent field states, it may perhaps not be surprising that in our scenario the detectors cannot harvest entanglement from the field: their delta couplings are very sudden and hence may introduce too much noise.

However, this may not be the only possible way to understand Theorem 3. In \cite{Pozas2017} it was shown (within perturbation theory) that gapless detectors cannot harvest vacuum entanglement at leading order. In light of this, another possible way to intuit why delta-coupled detectors cannot harvest field entanglement is to argue that during the interaction of each detector with the field, the infinite intensity of the detector's interaction Hamiltonian overwhelms its free Hamiltonian, and hence results in a complete lack of free detector dynamics~\cite{Pozas2017}. It is possible that the ability of detectors to evolve freely is crucial to them being able to harvest field entanglement. Indeed this is the case when we vary a detector's energy gap: a non-zero gap is necessary for non-trivial free dynamics, as well as for vacuum entanglement harvesting~\cite{Pozas2017}. 

Elucidating whether the inability of delta-coupled detectors to harvest entanglement is due to the noisy nature of their coupling or to their lack of free dynamics, is yet to be determined. In many cases these two pictures are related, but one can think of scenarios where the two are distinct. This could constitute an interesting question to be explored in future work.

%A promising approach to this problem is to consider two detectors that each delta-couple to the field on more than one occasion, since this would maintain the `noisy' aspect of the delta-switching, while allowing for non-trivial free evolution of each detector between its multiple field interaction times. This will be explored in future work.

\section{Conclusions}
\label{sec:conclusions}

We have considered the interactions of one and two Unruh-DeWitt particle detectors with an arbitrary coherent state of a scalar field. The detectors, initialized to their ground states, couple to the field by means of a Dirac-delta (short and intense) switching function, while no assumptions are made on their spatial profiles or internal energy gaps. We have obtained three main non-perturbative results regarding the evolved state of the detector/detectors:

First, we have shown that for a single detector interacting with the field, the eigenvalues of the time evolved density matrix $\rhoa$ of the detector are the same regardless of which coherent state the field was in prior to the interaction. This result therefore shows that any property of the evolved detector that is determined by the spectrum of its density matrix is independent of which coherent field state the detector interacted with. For instance the von Neumann entropy of the detector (which is also the entanglement entropy of the detector-field system) is such a property: the detector will get equally entangled with the field regardless of the coherent state of the field that it interacts with (such as the vacuum).

Second, we obtained an analogous result for two detectors interacting with the field. Namely, we found that the eigenvalues of the evolved two-detector density matrix $\rhoab$ are independent of which coherent field state the detectors interact with. Furthermore, the eigenvalues of the partially transposed two-detector density matrix $\rhoabpt$ are also independent of the field's coherent amplitude. Hence, any property of the evolved two-detector state that is determined solely by these eigenvalues is independent of what coherent state the field was in prior to the interaction.

This has strong consequences for entanglement harvesting from the field to the detectors:  The negativity of the two detector state (a measure of the entanglement between the detector pair) only depends on the eigenvalues of the partially transposed density matrix $\rhoabpt$. Therefore, the two detectors will harvest the same amount of entanglement from any coherent field state as they would from the vacuum. 

The fact that the eigenvalues of $\rhoa$, $\rhoab$, and $\rhoabpt$ are all independent of the coherent amplitude of the field, non-perturbatively for a delta switching and perturbatively for any switching~\cite{Simidzija2017b}, suggests that, at the very least, the spectra of these matrices may in general be extremely insensitive to the coherent amplitude of the field. However whether it is the case that these spectra are fully independent of the field's coherent amplitude in a non-perturbative regime for any arbitrary switching remains to be proved.

Finally, we have shown that the eigenvalues of the evolved two-detector density matrix, and the eigenvalues of its partial transpose, are equal. Interestingly, this shows that the eigenvalues of the partially transposed density matrix are non-negative, and therefore that the negativity of the evolved two-detector state is zero.

This allowed us to make a remarkable claim:  two detectors that couple to a coherent state of the field through delta-switching functions \textit{cannot} harvest any entanglement at all from the field. In particular this is true for entanglement harvesting from the vacuum, which is a particular coherent state.

As a final comment, we have offered two physical interpretations of this last result. The first suggests that delta-coupled detectors are unable to harvest entanglement from the field because their intense switchings introduce too much local noise that overcomes any non-local correlations that may arise between the detector pair~\cite{Reznik2005,Pozas2015}. Alternatively, it may be the case that the detectors cannot harvest entanglement because their delta-couplings result in a complete lack of free dynamics, and the ability of the detectors to evolve freely may be very important to their ability to harvest field entanglement~\cite{Pozas2017}. We plan to explore these two hypotheses in future work.%We have proposed a scenario that would, in principle, allow one to distinguish between these two phenomena, in which each of the two detectors is allowed to interact with the field through a delta-coupling, but at more than one instant in time. This would allow for non-trivial free evolution of each detector, while maintaining the noise inducing delta switching functions. 

\section*{Acknowledgments}
	
The work of P. S. and E. M.-M. is supported by the Natural Sciences and Engineering Research Council of Canada through the USRA and Discovery programs. E. M.-M. also gratefully acknowledges the funding of his Ontario Early Research Award.

\onecolumngrid
\appendix

\section{Calculations of inner products between two coherent states}
\label{Appendix:coherent_inner_product}

\subsection{Inner product of an arbitrary coherent state with the vacuum state}

Let us introduce some standard notation that will be useful throughout this appendix. Consider a string of bosonic creation and annihilation operators $\hat{A}_1\hat{A}_2\dots\hat{A}_m$. We define the \textit{normal-ordered} string $\normord{\hat{A}_1\hat{A}_2\dots\hat{A}_m}$ to be the permuted version of the original string such that all of the creation operators appear before all of the annihilation operators. Next we define the \textit{contraction} of two operators $\hat{A}$ and $\hat{B}$ to be:
\begin{equation}
    \contraction{}{\hat{A}}{}{
    \hat{B}}
    \hat{A}\hat{B}
    =
    \hat{A}\hat{B}-
    \normord{\hat{A}\hat{B}}.
\end{equation}
It is straightforward to show that
\begin{equation}
\label{eq:contractions1}
    \contraction{}{\ah}{{}_{\bm{k}}}{\ah}
    \a{k}\a{k'}
    =
    \contraction{}{\ah}{{}_{\bm{k}}^\dagger}{\ah}
    \ad{k}\ad{k'}
    =
    \contraction{}{\ah}{{}_{\bm{k}}^\dagger}{\ah}
    \ad{k}\a{k'}
    =
    0,
    \qquad
    \contraction{}{\ah}{{}_{\bm{k}}}{\ah}
    \a{k}\ad{k'}
    =
    \delta(\bm{k}-\bm{k'}).
\end{equation}
Finally, for ease of notation we will denote $\an{\bm{k}_j}$ by $\an{j}$. We can now proceed to the following lemma:
\vspace{2mm}
\\
\textbf{Lemma:} Let $\alpha_j$ be any scalars. Then,
\begin{equation}
    \bra{0}
    \prod_{j=1}^m
    \left(\alpha_j\adn{j}
    -\alpha_j^*\an{j}\right)
    \ket{0}
    =
    \begin{cases}
    (-1)^\frac{m}{2}\!\!
    \sum\limits_{\sigma\in S} 
    \underbrace{
    \left[\delta(\bm{k}_{\sigma(1)}-
    \bm{k}_{\sigma(2)})
    \alpha_{\sigma(1)}
    \alpha_{\sigma(2)}^*
    \right]
    \ldots
    \left[\delta(\bm{k}_{\sigma(m-1)}-
    \bm{k}_{\sigma(m)})
    \alpha_{\sigma(m-1)}
    \alpha_{\sigma(m)}^*
    \right]}_{m/2 \text{ factors}},& \!m \text{ even}
    \\
    0,& \!m \text{ odd},
    \end{cases}
\end{equation}
where $S$ is the set of all permutations $\sigma$ of $\{1,2,\ldots,m\}$ such that i) no two terms in the sum are equal, and ii) $\sigma(2j-1)\le\sigma(2j)$ for all $j$.
\\
\textit{Proof:} The proof for odd $m$ is obvious, since the vacuum expectation of an odd number of creation and annihilation operators is always zero. Therefore we can assume that $m$ is even. Now let us define the two-element sets $S_1, S_2, \ldots, S_m$ as \mbox{$S_j\coloneqq\{-\alpha_j^*\an{j},\alpha_j\adn{j}\}$}. Then
\begin{equation}
    \prod_{j=1}^m
    \left(\alpha_j\adn{j}
    -\alpha_j^*\an{j}\right)
    =
    \sum_{\hat{A}_j\in S_j}
    \hat{A}_1\hat{A}_2\ldots\hat{A}_m.
\end{equation}
By Wick's theorem (see, e.g.~\cite{Peskin1996}), this can be expressed as
\begin{align}
    \prod_{j=1}^m
    \left(\alpha_j\adn{j}
    -\alpha_j^*\an{j}\right)
    =
    \sum_{\hat{A}_j\in S_j}
    \Bigg[&
    \normord{\hat{A}_1\hat{A}_2\hat{A}_3\hat{A}_4\ldots\hat{A}_{m-1}\hat{A}_m}
    \notag\\
    &+
    \left(
    \normord{
    \contraction{}{\hat{A}}{{}_1}{\hat{A}}
    \hat{A}_1\hat{A}_2\hat{A}_3\hat{A}_4\ldots\hat{A}_{m-1}\hat{A}_m}
    +\text{ all other 1-pair contractions}
    \right)
    \notag\\
    &+
    \left(
    \normord{
    \contraction{}{\hat{A}}{{}_1}{\hat{A}}
    \contraction{\hat{A}_1\hat{A}_2}{\hat{A}}{{}_3}{\hat{A}}
    \hat{A}_1\hat{A}_2\hat{A}_3\hat{A}_4\ldots\hat{A}_{m-1}\hat{A}_m}
    +\text{ all other 2-pair contractions}
    \right)
    \notag\\
    &+\ldots
    \notag\\
    &+
    \left(
    \normord{
    \contraction{}{\hat{A}}{{}_1}{\hat{A}}
    \contraction{\hat{A}_1\hat{A}_2}{\hat{A}}{{}_3}{\hat{A}}
    \contraction{\hat{A}_1\hat{A}_2\hat{A}_3\hat{A}_4\ldots}{\hat{A}}{{}_{m-1}}{\hat{A}}
    \hat{A}_1\hat{A}_2\hat{A}_3\hat{A}_4\ldots\hat{A}_{m-1}\hat{A}_m}
    +\text{ all other $m/2$-pair contractions}
    \right)
    \Bigg].
\end{align}
Recall that $\ah_j\ket{0}=\bra{0}\ah_j^\dagger=0$. Therefore, since all of the terms on the right-hand side are normal-ordered, we find that after taking the vacuum expectation the terms which contain uncontracted operators will vanish. Therefore we obtain
\begin{equation}
    \bra{0}
    \prod_{j=1}^m
    \left(\alpha_j\adn{j}
    -\alpha_j^*\an{j}\right)
    \ket{0}
    =
    \sum_{\hat{A}_j\in S_j}
    \bra{0}
    \left(
    \normord{
    \contraction{}{\hat{A}}{{}_1}{\hat{A}}
    \contraction{\hat{A}_1\hat{A}_2}{\hat{A}}{{}_3}{\hat{A}}
    \contraction{\hat{A}_1\hat{A}_2\hat{A}_3\hat{A}_4\ldots}{\hat{A}}{{}_{m-1}}{\hat{A}}
    \hat{A}_1\hat{A}_2\hat{A}_3\hat{A}_4\ldots\hat{A}_{m-1}\hat{A}_m}
    +\text{ all other $m/2$-pair contractions}
    \right)
    \ket{0}.
\end{equation}
First, notice that all terms on the right hand side of this expression are distinct. Also, because of~\eqref{eq:contractions1}, a term is non-zero if and only if it can be written as
\begin{equation}
\label{eq:contractions2}
    \bra{0}
    \normord{
    \contraction{}{\hat{A}}{{}_{\sigma(1)}}{\hat{A}}
    \contraction{\hat{A}_{\sigma(1)}
    \hat{A}_{\sigma(2)}}{\hat{A}}{{}_{\sigma(3)}}{\hat{A}}
    \contraction{\hat{A}_{\sigma(1)}
    \hat{A}_{\sigma(2)}
    \hat{A}_{\sigma(3)}
    \hat{A}_{\sigma(4)}
    \ldots}{\hat{A}}{{}_{\sigma(m-1)}}{\hat{A}}
    \hat{A}_{\sigma(1)}
    \hat{A}_{\sigma(2)}
    \hat{A}_{\sigma(3)}
    \hat{A}_{\sigma(4)}
    \ldots
    \hat{A}_{\sigma(m-1)}
    \hat{A}_{\sigma(m)}
    }
    \ket{0},
\end{equation}
where $\sigma$ is a permutation of $\{1,2,\ldots,m\}$ such that i) \mbox{$\hat{A}_{\sigma(2j-1)}=-\alpha_j^*\ah_j$} and \mbox{$\hat{A}_{\sigma(2j)}=\alpha_j\ah_j^\dagger$} for all $j$, and ii) \mbox{$\sigma(2j-1)\le \sigma(2j)$} for all $j$. Using~\eqref{eq:contractions1}, expression \eqref{eq:contractions2} evaluates to
\begin{equation}
    (-1)^\frac{m}{2}
    \underbrace{
    \left[\delta(\bm{k}_{\sigma(1)}-
    \bm{k}_{\sigma(2)})
    \alpha_{\sigma(1)}
    \alpha_{\sigma(2)}^*
    \right]
    \ldots
    \left[\delta(\bm{k}_{\sigma(m-1)}-
    \bm{k}_{\sigma(m)})
    \alpha_{\sigma(m-1)}
    \alpha_{\sigma(m)}^*
    \right]}_{m/2 \text{ factors}},
\end{equation}
which completes the proof of the lemma.

We can now turn to evaluating the expression $\bra{0}\bnu\rangle$, where $\ket{\bnu}$ is an arbitrary coherent state. By the definition~\eqref{eq:alpha}, this can be written as the ``vacuum expectation" of a displacement operator, $\bra{0}\bnu\rangle= \bra{0}\hat{D}_{\bnu} \ket{0}$. (Note that we are slightly abusing the term ``vacuum expectation" since $\hat{D}_{\bnu}$ is not a self-adjoint operator, and hence there is no probabilistic interpretation of the expression $\bra{0}\hat{D}_{\bnu}\ket{0}$.) Using the definition of $\hat{D}_{\bnu}$ in~\eqref{eq:alpha}, this becomes
\begin{equation}
    \bra{0}\bnu\rangle
    =
    \bra{0}
    \exp\!
	\left(\int\!\!\d[n]{\bm{k}}\!\!\left[\bnu\ad{k}
	\!-\!
	\beta_\nu^*(\bm k)\a{k}\right]\!\right)
	\ket{0}.
\end{equation}
Expanding the exponential in a Taylor series, we obtain
\begin{equation}
    \bra{0}\bnu\rangle
    =
    \sum_{m=0}^\infty
    \frac{1}{m!}
    \left(
    \int\d[n]{\bm{k}_1}
    \ldots
    \int\d[n]{\bm{k}_m}
    \right)
    \bra{0}
    \prod_{j=1}^m
    \left(\beta_\nu(\bm{k}_j)\adn{j}
    -\beta_\nu^*(\bm{k}_j)\an{j}\right)
    \ket{0},
\end{equation}
where it is understood that the $m=0$ term is equal to $1$. Using the lemma, this becomes
\begin{align}
    \bra{0}\bnu\rangle
    =
    \sum_{m \text{ even}}\!\!
    \frac{(-1)^\frac{m}{2}}{m!}
    \left(
    \int\d[n]{\bm{k}_1}
    \ldots
    \int\d[n]{\bm{k}_m}
    \right)
    \!\sum_{\sigma\in S}
    &
    \left[\delta(\bm{k}_{\sigma(1)}-
    \bm{k}_{\sigma(2)})
    \beta_\nu(\bm{k}_{\sigma(1)})
    \beta_\nu^*(\bm{k}_{\sigma(2)})
    \right]
    \ldots
    \notag\\
    &\times
    \left[\delta(\bm{k}_{\sigma(m-1)}-
    \bm{k}_{\sigma(m)})
    \beta_\nu(\bm{k}_{\sigma(m-1)})
    \beta_\nu^*(\bm{k}_{\sigma(m)})
    \right],
\end{align}
where $S$ is the set of all permutations $\sigma$ of $\{1,2,\ldots,m\}$ such that i) no two terms in the sum are equal, and ii) $\sigma(2j-1)\le\sigma(2j)$ for all $j$. After relabelling the integration variables and performing the delta integrations, this becomes
\begin{equation}
    \bra{0}\bnu\rangle
    =
    \sum_{m \text{ even}}\!\!
    \frac{(-1)^\frac{m}{2}}{m!}
    \sum_{\sigma\in S}
    \left(
    \int\d[n]{\bm k}
    \left|
    \bnu
    \right|^2
    \right)^\frac{m}{2}.
\end{equation}
A simple inductive argument shows that, for a given $m$, there are $(m-1)!!$ elements in the set $S$, where the \textit{double factorial} is defined as
\begin{equation}
    n!!\coloneqq
    \begin{cases}
    (n)(n-2)(n-4)\ldots(4)(2),& n \text{ even}
    \\
    (n)(n-2)(n-4)\ldots(3)(1),& n \text{ odd}
    \end{cases},
\end{equation}
and $(-1)!!$ and $0!!$ are both defined to be 1. Hence
\begin{align}
    \bra{0}\bnu\rangle
    &=
    \sum_{m \text{ even}}\!\!
    \frac{(-1)^\frac{m}{2}(m-1)!!}{m!}
    \left(
    \int\d[n]{\bm k}
    \left|
    \bnu
    \right|^2
    \right)^\frac{m}{2}
    \notag\\
    &=
    \sum_{s=0}^\infty
    \frac{(-1)^s(2s-1)!!}{(2s)!}
    \left(
    \int\d[n]{\bm k}
    \left|
    \bnu
    \right|^2
    \right)^s,
\end{align}
where in the second line we have defined $s\coloneqq m/2$. Using the fact that
\begin{equation}
    \frac{(2s-1)!!}{(2s)!}
    =
    \frac{(2s-1)(2s-3)\ldots(3)(1)}{(2s)(2s-1)\dots(2)(1)}
    =
    \frac{1}{(2s)(2s-2)\ldots(4)(2)}
    =
    \frac{1}{2^s s!},
\end{equation}
we obtain
\begin{align}
\label{eq:inner_prod1}
    \bra{0}\bnu\rangle
    &=
    \sum_{s=0}^\infty
    \frac{1}{s!}
    \left(-\frac{1}{2}
    \int\d[n]{\bm k}
    \left|
    \bnu
    \right|^2
    \right)^s
    \notag\\
    &=
    \exp
    \left(
    -\frac{1}{2}
    \int\d[n]{\bm k}
    \left|
    \bnu
    \right|^2
    \right).
\end{align}

\subsection{Inner product of two arbitrary coherent states}

We will now evaluate the expression $\langle\bb\ket{\ba}$, where $\ket{\bb}$ and $\ket{\ba}$ are arbitrary coherent states. Using our definition of a coherent state~\eqref{eq:alpha}, we can express this as the ``vacuum expectation" of a product of displacement operators:
\begin{equation}
    \langle\bb\ket{\ba}
    =
    \bra{0}
    \hat{D}_{\bb}^\dagger
    \hat{D}_{\ba}
    \ket{0}
    =
    \bra{0}
    \hat{D}_{-\bb}
    \hat{D}_{\ba}
    \ket{0},
\end{equation}
where the last equality is easily seen from the definition~\eqref{eq:alpha} of a displacement operator. Making use of the commutator
\begin{align}
    \left[
    \int\d[n]{\bm{k}}
    \left(-\bb\ad{k}+
    \beta_\textsc{a}^*(\bm{k})
    \a{k}\right),
    \int\d[n]{\bm{k}'}
    \left(\beta_\textsc{a}(\bm{k}')\ad{k'}-
    \beta_\textsc{a}^*(\bm{k}')\a{k'}\right)
    \right]
    =
    \int\d[n]{\bm k}
    \left(
     \beta_\textsc{b}^*(\bm{k})\ba-
    \bb \beta_\textsc{a}^*(\bm{k})
    \right)\mathds{1}_{\hat{\phi}},
\end{align}
and the Baker-Campbell-Hausdorff formula (see, e.g.~\cite{Truax1988}), we find that
\begin{equation}
    \langle\bb\ket{\ba}
    =
    \bra{0}
    \hat{D}_{\ba-\bb}
    \ket{0}
    \exp 
    \left[
    \frac{1}{2}
    \int\d[n]{\bm k}
    \left(
    \beta_\textsc{b}^*(\bm{k})\ba-
    \bb \beta_\textsc{a}^*(\bm{k})
    \right)
    \right].
\end{equation}
Using expression~\eqref{eq:inner_prod1} we obtain
\begin{align}
    \langle\bb\ket{\ba}
    &=
    \exp 
    \left[
    -\frac{1}{2}
    \int\d[n]{\bm k}
    \left|
    \ba-\bb
    \right|^2
    \right]
    \exp 
    \left[
    \frac{1}{2}
    \int\d[n]{\bm k}
    \left(
    \beta_\textsc{b}^*(\bm{k})\ba-
    \bb \beta_\textsc{a}^*(\bm{k})
    \right)
    \right]
    \notag\\
    &=
    \exp 
    \left[
    -\frac{1}{2}
    \int\d[n]{\bm k}
    \left(
    |\ba|^2+
    |\bb|^2-
    2 \ba\beta_\textsc{b}^*(\bm k)
    \right)
    \right].
\end{align}

\section{Comparisons of density matrices in the perturbative regime}
\label{Appendix:pert}

\subsection{One-detector density matrix}
\label{Appendix:pert1}

Expanding $\rhoa$ in~\eqref{eq:rho_a_final} to $\mathcal{O}(\lambda_\textsc{a}^2)$ gives the expression
\begin{equation}
    \label{eq:rho_a_pert1}
    \rhoa=
    \begin{pmatrix}
    1+\bra{0}\Ya^2\ket{0}-\ca^2 &
    -\ii\ca \\
    \ii\ca & -\bra{0}\Ya^2\ket{0}+\ca^2
    \end{pmatrix}
    +\mathcal{O}(\lambda_\textsc{a}^3).
\end{equation}
Recall that we are working in the basis~\eqref{eq:B2} of the Hilbert space of detector A. Written in this basis, equation (46) for $\rhoa$ in~\cite{Simidzija2017b} takes the form
\begin{equation}
\label{eq:rho_a_pert2}
	\hat{\rho}_\textsc{a}=
	\begin{pmatrix}
	1-\mathcal{L}_{\textsc{aa}}-
	\bar{\mathcal{L}}_{\textsc{aa}} & \bar{L}_\textsc{a}^* e^{\ii\Omega_\textsc{a}t_\textsc{a}}\\
	\bar{L}_\textsc{a} e^{-\ii\Omega_\textsc{a}t_\textsc{a}} & \mathcal{L}_{\textsc{aa}}+
	\bar{\mathcal{L}}_{\textsc{aa}}
	\end{pmatrix}
	+\mathcal{O}(\lambda_\textsc{a}^3).
\end{equation}
The terms in the above matrix are
\begin{align}
\label{eq:Laa}
    \mathcal{L}_{\textsc{aa}}
	&=
    \int\d[n]{\bm{k}}
    L_\textsc{a}(\bm{k})
    L_\textsc{a}(\bm{k})^*,
    \\
    \bar{\mathcal{L}}_{\textsc{aa}}
	&=
	\bar{L}_\textsc{a}\bar{L}_\textsc{a}^*,
	\\
	\bar{L}_\textsc{a}
	&=
	-\ii\lambda_\textsc{a}
	\int_{-\infty}^{\infty}\!\!\!\!\dif t			\chi_\textsc{a}(t)e^{\ii\Omega_\textsc{a} t}V(\bm{x}_\textsc{a},t),
\end{align}
with $L_\textsc{a}(\bm k)$ and $V(\bm{x}_\textsc{a}, t)$ given by
\begin{align}
    L_\textsc{a}(\bm{k})
	&=
	\lambda_\textsc{a}
	\frac{e^{-\ii\bm{k}\cdot\bm{x}_\textsc{a}}
	\tilde{F}_\textsc{a}(\bm{k})^*}{\sqrt{2|\bm{k}|}}
	\int_{-\infty}^{\infty}\!\!\!\!\!\dif t \,\chi_\textsc{a}(t)
	e^{\ii(|\bm{k}|+\Omega_\textsc{a})t},
	\\
	V(\bm{x}_\textsc{a},t)
	&=
	\int\!\!
	\frac{\d[n]{\bm{k}}}{\sqrt{2|\bm k|}}
	\left(
	F_\textsc{a}(\bm k)
	\alpha(\bm k)
	e^{-\ii(|\bm k|t-\bm k\cdot \bm x_\textsc{a})}
	+\text{c.c.}
	\right),
\end{align}
and where $\tilde{F}_\textsc{a}(\bm{k})$ is the Fourier transform of the detector's smearing function $F_\textsc{a}(\bm{x})$,
\begin{equation}
	\tilde{F}_\textsc{a}(\bm{k})
	\coloneqq
	\frac{1}{\sqrt{(2\pi)^n}}\int\d[n]{\bm{x}}
	F_\textsc{a}(\bm{x})e^{\ii\bm{k}\cdot\bm{x}}.
\end{equation}
We would like to show that if we set the detector's switching function $\chi_\textsc{a}(t)$ to a delta function, as in~\eqref{eq:chi_delta_a}, then the matrix~\eqref{eq:rho_a_pert2} reduces to the matrix~\eqref{eq:rho_a_pert1}. Using the definition~\eqref{eq:c_a} of $\ca$, it is immediately obvious that (for delta switching)
\begin{equation}
    \bar{L}_\textsc{a}
    =
    \ii\ca e^{\ii\Omega_\textsc{a}t_\textsc{a}}.
\end{equation}
From this we also obtain that
\begin{equation}
    \bar{\mathcal{L}}_\textsc{aa}=\ca^2.
\end{equation}
Hence to show that~\eqref{eq:rho_a_pert1} is equal to~\eqref{eq:rho_a_pert2} for delta detector switching, all that is left to show is that $-\mathcal{L}_\textsc{aa}$ is equal to $\bra{0}\Ya^2\ket{0}$. But this is true since (using the definition~\eqref{eq:Ya} for $\Ya$)
\begin{align}
    \bra{0}\Ya^2\ket{0}
    &=
    -\lambda_\textsc{a}^2
    \eta_\textsc{a}^2
    \!\int\!\!\d[n]{\bm x}\!\!\!
    \int\!\!\d[n]{\bm x'}
    F_\textsc{a}(\bm{x}-\bm{x}_\textsc{a})
    F_\textsc{a}(\bm{x'}-\bm{x}_\textsc{a})
    \bra{0}
    \hat{\phi}(\bm{x},t_\textsc{a})
    \hat{\phi}(\bm{x'},t_\textsc{a})
    \ket{0}
    \notag\\
    &=
    -\lambda_\textsc{a}^2
    \eta_\textsc{a}^2
    \!\int\!\!\d[n]{\bm x}\!\!\!
    \int\!\!\d[n]{\bm x'}
    F_\textsc{a}(\bm{x}-\bm{x}_\textsc{a})
    F_\textsc{a}(\bm{x'}-\bm{x}_\textsc{a})
    \!\int\!\!
    \d[n]{\bm k}
    \frac{e^{-\ii(|\bm{k}|t_\textsc{a}-\bm{k}\cdot\bm{x})}}{\sqrt{2(2\pi)^n|\bm{k}|}}
    \int\!\!
    \d[n]{\bm k'}
    \frac{e^{\ii(|\bm{k'}|t_\textsc{a}-\bm{k'}\cdot\bm{x'})}}{\sqrt{2(2\pi)^n|\bm{k'}|}}
    \bra{0}\a{k}\ad{k'}\ket{0}
    \notag\\
    &=
    -\lambda_\textsc{a}^2
    \eta_\textsc{a}^2
    \!\int\!\!\d[n]{\bm x}\!\!\!
    \int\!\!\d[n]{\bm x'}
    F_\textsc{a}(\bm{x}-\bm{x}_\textsc{a})
    F_\textsc{a}(\bm{x'}-\bm{x}_\textsc{a})
    \!\int\!\!
    \d[n]{\bm k}
    \frac{e^{\ii\bm{k}\cdot(\bm{x}-\bm{x'})}}{2(2\pi)^n|\bm{k}|}
    \notag\\
    &=
    -\frac{\lambda_\textsc{a}^2
    \eta_\textsc{a}^2}{2}
    \!\int\!\!
    \frac{\d[n]{\bm k}}{|\bm{k}|}
    \left| 
    \tilde{F}_\textsc{a}(\bm k)
    \right|^2,
\end{align}
and since (from the definition~\eqref{eq:Laa} of $\mathcal{L}_\textsc{aa}$)
\begin{align}
    \mathcal{L}_\textsc{aa}
    &=
    \frac{\lambda_\textsc{a}^2
    \eta_\textsc{a}^2}{2}
    \!\int\!\!
    \frac{\d[n]{\bm k}}{|\bm{k}|}
    \left| 
    \tilde{F}_\textsc{a}(\bm k)
    \right|^2.
\end{align}
Hence our expression for the time-evolved one detector density matrix $\rhoa$ is consistent with that obtained perturbatively in~\cite{Simidzija2017b}.

\subsection{Two-detector density matrix}
\label{Appendix:pert2}

Expanding $\rhoab$ in~\eqref{eq:rhoab} to $\mathcal{O}(\lambda_\textsc{a}^i\lambda_\textsc{b}^j)$ with $(i,j)\in\{(0,2),(1,1),(2,0)\}$, which we simply denote as $\mathcal{O}(\lambda_\nu^2)$, we obtain
\begin{equation}
\label{eq:rho_ab_pert1}
    \rhoab
    =
\begin{pmatrix}
    1+\bra{0}\Ya^2\ket{0}+\bra{0}\Yb^2\ket{0}-\ca^2-\cb^2 &
    -\ii\cb & -\ii\ca & \bra{0}\Yb\Ya\ket{0}^*-\ca\cb
    \\
    \ii\cb & -\bra{0}\Yb^2\ket{0} +\cb^2 & -\bra{0}\Yb\Ya\ket{0}^* +\ca\cb & 0
    \\
    \ii\ca & -\bra{0}\Yb\Ya\ket{0} +\ca\cb & -\bra{0}\Ya^2\ket{0} +\ca^2 & 0
    \\
    \bra{0}\Yb\Ya\ket{0}-\ca\cb & 0 & 0 & 0
\end{pmatrix}
+\mathcal{O}(\lambda_\nu^3).
\end{equation}
Recall that we are working in the basis~\eqref{eq:B2_2} of the two-detector Hilbert space. Written in this basis, equation (55) for $\rhoab$ in~\cite{Simidzija2017b} takes the form
\begin{align}
\label{eq:rho_ab_pert2}
	\rhoab
	=&
	\begin{pmatrix}
		1-\mathcal{L}_\textsc{aa}-\mathcal{L}_\textsc{bb} -\bar{\mathcal{L}}_\textsc{aa}-\bar{\mathcal{L}}_\textsc{bb} & \bar{L}_\textsc{b}^* e^{\ii\Omega_\textsc{b}t_\textsc{b}} & \bar{L}_\textsc{a}^* e^{\ii\Omega_\textsc{a}t_\textsc{a}} & \left(\mathcal{M}^* +\bar{\mathcal{M}}^*\right) e^{\ii\Omega_\textsc{a}t_\textsc{a}} e^{\ii\Omega_\textsc{b}t_\textsc{b}}
		\\
		\bar{L}_\textsc{b} e^{-\ii\Omega_\textsc{b}t_\textsc{b}} & \mathcal{L}_\textsc{bb} +\bar{\mathcal{L}}_\textsc{bb} & \left(\mathcal{L}_\textsc{ab}^* +\bar{\mathcal{L}}_\textsc{ab}^*\right) e^{\ii\Omega_\textsc{a}t_\textsc{a}} e^{-\ii\Omega_\textsc{b}t_\textsc{b}} & 0 \\
		\bar{L}_\textsc{a} e^{-\ii\Omega_\textsc{a}t_\textsc{a}} & \left(\mathcal{L}_\textsc{ab} +\bar{\mathcal{L}}_\textsc{ab}\right) e^{-\ii\Omega_\textsc{a}t_\textsc{a}} e^{\ii\Omega_\textsc{b}t_\textsc{b}}& \mathcal{L}_\textsc{aa} +\bar{\mathcal{L}}_\textsc{aa} & 0 \\
		\left(\mathcal{M} +\bar{\mathcal{M}}\right) e^{-\ii\Omega_\textsc{a}t_\textsc{a}} e^{-\ii\Omega_\textsc{b}t_\textsc{b}} & 0 & 0 & 0
	\end{pmatrix}
	\notag\\
	&+\mathcal{O}(\lambda_\nu^3).
\end{align}
The terms in the above matrix are
\begin{align}
\label{eq:L_mu_nu}
	\mathcal{L}_{\mu\nu}
	&=
	\int\d[n]{\bm{k}}L_\mu(\bm{k})L_\nu(\bm{k})^*,
	\\
\label{eq:M}
	\mathcal{M}
	&=
	\int\d[n]{\bm{k}}M(\bm{k}),
	\\
\label{eq:L_mu_nu_bar}
	\bar{\mathcal{L}}_{\mu\nu}
	&=
	\bar{L}_\mu\bar{L}_\nu^*,
	\\
\label{eq:M_bar}
	\bar{\mathcal{M}}
	&=
	\bar{L}_\textsc{a}
	\bar{L}_\textsc{b},
	\\
\label{eq:L_nu_bar}
	\bar{L}_\nu
	&=
	-\ii\lambda_\nu
	\int_{-\infty}^{\infty}\!\!\!\!\dif t			\chi_\nu(t)e^{\ii\Omega_\nu t}V(\bm{x}_\nu,t),
\end{align}
with $L_\nu(\bm{k})$, $M(\bm{k})$, and $V(\bm{x}_\nu,t)$ given by
\begin{align}
	L_\nu(\bm{k})
	&=
	\lambda_\nu\frac{e^{-\ii\bm{k}\cdot\bm{x}_\nu}
	\tilde{F}_\nu(\bm{k})^*}{\sqrt{2|\bm{k}|}}
	\int_{-\infty}^{\infty}\!\!\!\!\dif t \chi_\nu(t)
	e^{\ii(|\bm{k}|+\Omega_\nu)t},
	\\
	\label{eq:M2}
	M(\bm{k})
	&=
	-\frac{\lambda_\textsc{a}\lambda_\textsc{b}}{2|\bm{k}|}
	\int_{-\infty}^{\infty}
	\!\!\!\dif t
	\int_{-\infty}^{t}
	\!\!\!\dif t'
	e^{-\ii|\bm{k}|(t-t')}
	\Big[
	\chi_\textsc{a}(t)
	\chi_\textsc{b}(t')
	e^{\ii(\Omega_\textsc{a}t
	+\Omega_\textsc{b}t')}
	e^{\ii\bm{k}\cdot(\bm{x}_\textsc{a}-\bm{x}_\textsc{b})}
	\tilde{F}_\textsc{a}(\bm k)
	\tilde{F}_\textsc{b}(\bm k)^*
	\notag\\
	&\hspace{15em}+
	\chi_\textsc{b}(t)
	\chi_\textsc{a}(t')
	e^{\ii(\Omega_\textsc{b}t
	+\Omega_\textsc{a}t')}
	e^{\ii\bm{k}\cdot(\bm{x}_\textsc{b}-\bm{x}_\textsc{a})}
	\tilde{F}_\textsc{b}(\bm k)
	\tilde{F}_\textsc{a}(\bm k)^*
	\Big],
	\\
	V(\bm{x}_\nu,t)
	&=
	\int\!\!
	\frac{\d[n]{\bm{k}}}{\sqrt{2|\bm k|}}
	\left(
	F_\nu(\bm k)
	\alpha(\bm k)
	e^{-\ii(|\bm k|t-\bm k\cdot \bm x_\nu)}
	+\text{c.c.}
	\right),
\end{align}
and where $\tilde{F}_\nu(\bm{k})$ is the Fourier transform of the smearing function $F_\nu(\bm{x})$ of detector $\nu$,
\begin{equation}
	\tilde{F}_\nu(\bm{k})
	\coloneqq
	\frac{1}{\sqrt{(2\pi)^n}}\int\d[n]{\bm{x}}
	F_\nu(\bm{x})e^{\ii\bm{k}\cdot\bm{x}}.
\end{equation}
We would like to show that if we set the detectors' switching functions $\chi_\nu(t)$ to a delta functions, as in~\eqref{eq:chi_delta_nu}, then the matrix~\eqref{eq:rho_ab_pert2} reduces to the matrix~\eqref{eq:rho_ab_pert1}. Using the definitions~\eqref{eq:c_a} of $\ca$ and~\eqref{eq:c_b} of $\cb$, it is immediately obvious that (for delta switching)
\begin{equation}
    \bar{L}_\nu
    =
    \ii\cnu e^{\ii\Omega_\nu t_\nu}.
\end{equation}
From this we also obtain that
\begin{align}
    \bar{\mathcal{L}}_{\mu\nu}
    &=
    \cmu \cnu
    e^{\ii\Omega_\mu t_\mu}
    e^{-\ii\Omega_\nu t_\nu},
    \\
    \bar{\mathcal{M}}
    &=
    -\ca \cb
    e^{\ii\Omega_\textsc{a} t_\textsc{a}}
    e^{\ii\Omega_\textsc{b} t_\textsc{b}},
\end{align}
where we used the fact that the $\cnu$ are real. In the previous section of this appendix, we showed that
\begin{equation}
    \mathcal{L}_\textsc{aa}
    =
    -\bra{0}\Ya^2\ket{0}.
\end{equation}
Analogously one can readily show that
\begin{equation}
    \mathcal{L}_\textsc{bb}
    =
    -\bra{0}\Yb^2\ket{0}.
\end{equation}
Hence in order to show that~\eqref{eq:rho_ab_pert1} is equal to~\eqref{eq:rho_ab_pert2} in the case of delta switching, all that is left is to show that \mbox{$\mathcal{M}=\bra{0}\Yb\Ya\ket{0}e^{\ii\Omega_\textsc{a}t_\textsc{a}}e^{\ii\Omega_\textsc{b}t_\textsc{b}}$} and that \mbox{$\mathcal{L}_\textsc{ab}=-\bra{0}\Yb\Ya\ket{0}e^{\ii\Omega_\textsc{a}t_\textsc{a}}e^{-\ii\Omega_\textsc{b}t_\textsc{b}}$}. But this is true since (using definition~\eqref{eq:Ya} for $\Ya$ and~\eqref{eq:Yb} for $\Yb$)
\begin{align}
    \bra{0}\Yb\Ya\ket{0}
    &=
    -\lambda_\textsc{a}
    \lambda_\textsc{b}
    \eta_\textsc{a}
    \eta_\textsc{b}
    \!\int\!\!\d[n]{\bm x}\!\!\!
    \int\!\!\d[n]{\bm x'}
    F_\textsc{b}(\bm{x}-\bm{x}_\textsc{b})
    F_\textsc{a}(\bm{x'}-\bm{x}_\textsc{a})
    \bra{0}
    \hat{\phi}(\bm{x},t_\textsc{b})
    \hat{\phi}(\bm{x'},t_\textsc{a})
    \ket{0}
    \notag\\
    &=
    -\lambda_\textsc{a}
    \lambda_\textsc{b}
    \eta_\textsc{a}
    \eta_\textsc{b}
    \!\int\!\!\d[n]{\bm x}\!\!\!
    \int\!\!\d[n]{\bm x'}
    F_\textsc{b}(\bm{x}-\bm{x}_\textsc{b})
    F_\textsc{a}(\bm{x'}-\bm{x}_\textsc{a})
    \!\int\!\!
    \d[n]{\bm k}
    \frac{e^{-\ii(|\bm{k}|t_\textsc{b}-\bm{k}\cdot\bm{x})}}{\sqrt{2(2\pi)^n|\bm{k}|}}
    \int\!\!
    \d[n]{\bm k'}
    \frac{e^{\ii(|\bm{k'}|t_\textsc{a}-\bm{k'}\cdot\bm{x'})}}{\sqrt{2(2\pi)^n|\bm{k'}|}}
    \bra{0}\a{k}\ad{k'}\ket{0}
    \notag\\
    &=
    -\lambda_\textsc{a}
    \lambda_\textsc{b}
    \eta_\textsc{a}
    \eta_\textsc{b}
    \!\int\!\!\d[n]{\bm x}\!\!\!
    \int\!\!\d[n]{\bm x'}
    F_\textsc{b}(\bm{x}-\bm{x}_\textsc{b})
    F_\textsc{a}(\bm{x'}-\bm{x}_\textsc{a})
    \!\int\!\!
    \d[n]{\bm k}
    \frac{ e^{-\ii(|\bm{k}|t_\textsc{b}-\bm{k}\cdot\bm{x})} e^{\ii(|\bm{k}|t_\textsc{a}-\bm{k}\cdot\bm{x'})}}{2(2\pi)^n|\bm{k}|}
    \notag\\
    &=
    -\frac{\lambda_\textsc{a}
    \lambda_\textsc{b}
    \eta_\textsc{a}
    \eta_\textsc{b}}{2}
    \!\int\!\!
    \frac{\d[n]{\bm k}}{|\bm{k}|}
    \tilde{F}_\textsc{a}(\bm k)^*
    \tilde{F}_\textsc{b}(\bm k)
    e^{\ii|\bm{k}|(t_\textsc{a}-t_\textsc{b})}
    e^{-\ii\bm{k}\cdot(\bm{x}_\textsc{a}-\bm{x}_\textsc{b})}
    ,
\end{align}
and since (from definition~\eqref{eq:M} of $\mathcal{M}$ and definition~\eqref{eq:L_mu_nu} of $\mathcal{L}_\textsc{ab}$)
\begin{align}
    \mathcal{M}
    &=
    -\frac{\lambda_\textsc{a}
    \lambda_\textsc{b}
    \eta_\textsc{a}
    \eta_\textsc{b}}{2}
    \!\int\!\!
    \frac{\d[n]{\bm k}}{|\bm{k}|}
    \tilde{F}_\textsc{a}(\bm k)^*
    \tilde{F}_\textsc{b}(\bm k)
    e^{\ii|\bm{k}|(t_\textsc{a}-t_\textsc{b})}
    e^{-\ii\bm{k}\cdot(\bm{x}_\textsc{a}-\bm{x}_\textsc{b})}
    e^{\ii\Omega_\textsc{a}t_\textsc{a}}
    e^{\ii\Omega_\textsc{b}t_\textsc{b}},
    \\
    \mathcal{L}_\textsc{ab}
    &=
    \frac{\lambda_\textsc{a}
    \lambda_\textsc{b}
    \eta_\textsc{a}
    \eta_\textsc{b}}{2}
    \!\int\!\!
    \frac{\d[n]{\bm k}}{|\bm{k}|}
    \tilde{F}_\textsc{a}(\bm k)^*
    \tilde{F}_\textsc{b}(\bm k)
    e^{\ii|\bm{k}|(t_\textsc{a}-t_\textsc{b})}
    e^{-\ii\bm{k}\cdot(\bm{x}_\textsc{a}-\bm{x}_\textsc{b})}
    e^{\ii\Omega_\textsc{a}t_\textsc{a}}
    e^{-\ii\Omega_\textsc{b}t_\textsc{b}}.
\end{align}
Hence our expression for the time-evolved two detector density matrix $\rhoab$ is consistent with that obtained perturbatively in~\cite{Simidzija2017b}. 

As a final comment, note that special care must be taken if the detectors' interaction times coincide (i.e. if $t_\textsc{a}=t_\textsc{b}$), since then we must evaluate nested integrals in~\eqref{eq:M2}, in which an integration limit coincides with where the delta switching functions of the detectors are centered. Without further assumptions on the switching functions, these integrals are ambiguously defined. However, we must recall that when we introduced the delta switching functions, we also made the assumption that they are the symmetrically taken limit of a symmetric regularization function. This ensures that the nested integrals are well defined even in the case of coinciding detectors, and it also ensures that the perturbative expansion of our non-perturbative results agrees with the perturbative results of~\cite{Simidzija2017b}. For a more detailed discussion of the importance of regularizations of instantaneous switching functions, see appendix D in~\cite{Pozas2017}.

\section{Explicit calculation of \texorpdfstring{$\hat{U}_2$}{}}
\label{Appendix:U2}

In the case of delta detector switching functions~\eqref{eq:chi_delta_nu}, the unitary $\hat{U}_2$ given in~\eqref{eq:U2_def} becomes
\begin{align}
    \hat{U}_2
    =
    \mathcal{T}
    \exp
    \left(
    -\ii 
    \left[
    \hat{H}_\textsc{i,a}^{(2)}(t_\textsc{a})
    +\hat{H}_\textsc{i,b}^{(2)}(t_\textsc{b})
    \right]
    \right).
\end{align}
Expanding the exponential as a power series, one obtains
\begin{align}
    \hat{U}_2
    &=
    \sum_{n=0}^\infty
    \frac{1}{n!}
    \mathcal{T}
    \left(
    \hat{H}_\textsc{i,a}^{(2)}(t_\textsc{a})
    +\hat{H}_\textsc{i,b}^{(2)}(t_\textsc{b})
    \right)^n.
\end{align}
Recalling that, wlog, we assumed $t_\textsc{a}\le t_\textsc{b}$, this becomes
\begin{align}
    \hat{U}_2
    &=
    \sum_{n=0}^\infty
    \frac{1}{n!}
    \sum_{m=0}^n
    \binom{n}{m}
    \left(
    \hat{H}_\textsc{i,b}^{(2)}(t_\textsc{b})
    \right)^m
    \left(
    \hat{H}_\textsc{i,a}^{(2)}(t_\textsc{a})
    \right)^{n-m}
    \notag\\
    &=
    \sum_{n=0}^\infty
    \sum_{m=0}^n
    \frac{1}{m!(n-m)!}
    \left(
    \hat{H}_\textsc{i,b}^{(2)}(t_\textsc{b})
    \right)^m
    \left(
    \hat{H}_\textsc{i,a}^{(2)}(t_\textsc{a})
    \right)^{n-m}.
\end{align}
We now define the following two sets:
\begin{align}
    S_1
    &\coloneqq
    \{
    (x,y)\in{\rm I\!R}^2|0\le y\le x
    \},
    \\
    S_2
    &\coloneqq
    \{
    (x,y)\in{\rm I\!R}^2|0\le x,0\le y
    \}.
\end{align}
Now we define the map $f:S_1\rightarrow S_2$ by
\begin{equation}
    f(x,y)\coloneqq (x-y,y).
\end{equation}
It is straightforward to show that $f$ is a bijection. Hence defining new summation indices $k$ and $l$ by \mbox{$(k,l)\coloneqq f(n,m)$} allows us to write $\hat{U}_2$ as
\begin{align}
    \hat{U}_2
    &=
    \sum_{(n,m)\in S_1}
    \frac{1}{m!(n-m)!}
    \left(
    \hat{H}_\textsc{i,b}^{(2)}(t_\textsc{b})
    \right)^m
    \left(
    \hat{H}_\textsc{i,a}^{(2)}(t_\textsc{a})
    \right)^{n-m}
    \notag\\
    &=
    \sum_{(k,l)\in S_2}
    \frac{1}{l!\,k!}
    \left(
    \hat{H}_\textsc{i,b}^{(2)}(t_\textsc{b})
    \right)^l
    \left(
    \hat{H}_\textsc{i,a}^{(2)}(t_\textsc{a})
    \right)^k
    \notag\\
    &=
    \sum_{l=0}^\infty
    \sum_{k=0}^\infty
    \frac{1}{l!\,k!}
    \left(
    \hat{H}_\textsc{i,b}^{(2)}(t_\textsc{b})
    \right)^l
    \left(
    \hat{H}_\textsc{i,a}^{(2)}(t_\textsc{a})
    \right)^k
    \notag\\
    &=
    \exp
    \left(-\ii 
    \hat{H}_\textsc{i,b}^{(2)}(t_\textsc{b})
    \right)
    \exp
    \left(-\ii 
    \hat{H}_\textsc{i,a}^{(2)}(t_\textsc{a})
    \right),
\end{align}
which proves the statement in~\eqref{eq:U2}.

\section{Explicit calculation of \texorpdfstring{$f_2^{(jklm)}$}{}}
\label{Appendix:f2}

Using the Baker-Campbell-Hausdorff (BCH) formula (see, e.g.~\cite{Truax1988}), the definition~\eqref{eq:X_jk} for $\hat{X}_{(lm)}$, and the commutators $\ca$ in~\eqref{eq:c_a} and $\cb$ in~\eqref{eq:c_b}, we obtain
\begin{align}
\label{eq:XD}
    \hat{X}_{(lm)}\disp
    =
    \frac{1}{4}\disp
    \Big[
    &\left(e^{\Yb}e^{\ii\cb}\right)
    \left(e^{\Ya}e^{\ii\ca}\right)
    + m
    \left(e^{\Yb}e^{\ii\cb}\right)
    \left(e^{-\Ya}e^{-\ii\ca}\right)
    \notag\\
    &+ l
    \left(e^{-\Yb}e^{-\ii\cb}\right)
    \left(e^{\Ya}e^{\ii\ca}\right)
    + ml
    \left(e^{-\Yb}e^{-\ii\cb}\right)
    \left(e^{-\Ya}e^{-\ii\ca}\right)
    \Big].
\end{align}
Taking the Hermitean adjoint and relabeling the indices gives
\begin{align}
\label{eq:DX}
    \disp^\dagger\hat{X}_{(jk)}^\dagger
    =
    \frac{1}{4}
    \Big[
    &\left(e^{-\Ya}e^{-\ii\ca}\right)
    \left(e^{-\Yb}e^{-\ii\cb}\right)
    + k
    \left(e^{\Ya}e^{\ii\ca}\right)
    \left(e^{-\Yb}e^{-\ii\cb}\right)
    \notag\\
    &+ j
    \left(e^{-\Ya}e^{-\ii\ca}\right)
    \left(e^{\Yb}e^{\ii\cb}\right)
    + jk
    \left(e^{\Ya}e^{\ii\ca}\right)
    \left(e^{\Yb}e^{\ii\cb}\right)
    \Big]
    \disp^\dagger.
\end{align}
Next, we define $\theta$ as in~\eqref{eq:theta} so that $[\Ya,\Yb]=\ii\theta\mathds{1}_{\hat{\phi}}$. Then, using the BCH formula, we find that
\begin{equation}
\label{eq:eYbeYa}
    e^{p\Ya}e^{q\Yb}
    =
    e^{(pq)\ii\theta}
    e^{q\Yb}e^{p\Ya},
\end{equation}
for any scalars $p$ and $q$. Finally, using~\eqref{eq:XD},~\eqref{eq:DX} and~\eqref{eq:eYbeYa}, we find that $f_2^{(jklm)}$, as defined in~\eqref{eq:f2_def}, evaluates to~\eqref{eq:f2}.

\section{An explicit proof of the inequalities \texorpdfstring{$E_{_{\text{AB},\nu}}^{{\text{\textbf{t}}}_{_\text{A}}} \ge 0$}{}}
\label{Appendix:eigenvalues}

We first make note of three useful lemmas:
\vspace{2mm}
\\
\textbf{Lemma 1:} Let $c$ be a complex number. Then $|\sinh(c)|\le \sinh(|c|)$.
\\
\textit{Proof:} The proof is straightforward if we use the Taylor series expansion for $\sinh$:
\begin{equation}
    \left|\sinh(c)\right|=
    \left|\sum_{n=0}^\infty 
    \frac{c^{2n+1}}{(2n+1)!}\right|
    \le
    \sum_{n=0}^\infty
    \frac{|c|^{2n+1}}{(2n+1)!}
    =\sinh\left(|c|\right).
\end{equation}
Note that the inequality is well-defined since both sums converge. This completes the proof.
\vspace{2mm}
\\
\textbf{Lemma 2:} Let $g(\bm k)$ and $h(\bm k)$ be real-valued, non-negative, continuous functions of $\bm{k}$. Then
\begin{equation}
    \int\d[n]{\bm k} g(\bm k)^2
    \int\d[n]{\bm k'} h(\bm k')^2
    \ge 
    \left(
    \int\d[n]{\bm k} g(\bm k) h(\bm k)
    \right)^2.
\end{equation}
\\
\textit{Proof:} The proof is given in appendix B of~\cite{Simidzija2017b}.
\vspace{2mm}
\\
\textbf{Lemma 3:} Let $x$ and $y$ be real numbers. Then $\sinh(a^2)\sinh(b^2)\ge\sinh^2(ab)$.
\\
\textit{Proof:} Suppose $a=0$ or $b=0$. Then equality holds trivially. Otherwise, $ab\neq 0$. Then, consider Euler's well known infinite product expansion of $\sinh(a)$:
\begin{equation}
    \frac{\sinh(a)}{a}
    =
    \prod_{n=1}^{\infty}
    \left(1+\frac{a^2}{\pi^2 n^2}
    \right).
\end{equation}
Then we see that the desired inequality holds if and only if
\begin{align}
    &\frac{\sinh(a^2)}{a^2}
    \frac{\sinh(b^2)}{b^2}
    \ge \frac{\sinh^2(ab)}{a^2 b^2}
    \\
    \iff &
    \prod_{n=1}^{\infty}
    \left(1+\frac{a^4}{\pi^2 n^2}
    \right)
    \prod_{n=1}^{\infty}
    \left(1+\frac{b^4}{\pi^2 n^2}
    \right)
    \ge
    \left[
    \prod_{n=1}^{\infty}
    \left(1+\frac{a^2 b^2}{\pi^2 n^2}
    \right)
    \right]^2.
\end{align}
Since all of the infinite products converge, we can combine the two products on the left hand side of the inequality, and the two on the right hand side. Hence the desired inequality holds if and only if
\begin{equation}
    \prod_{n=1}^{\infty}
    \left(1+\frac{a^4+b^4}{\pi^2 n^2} +\frac{a^4 b^4}{\pi^4 n^4}
    \right)
    \ge
    \prod_{n=1}^{\infty}
    \left(1+\frac{2 a^2 b^2}{\pi^2 n^2} +\frac{a^4 b^4}{\pi^4 n^4}
    \right).
\end{equation}
However this always holds since $(a^2-b^2)^2\ge 0$ implies that $a^4+b^4\ge 2a^2b^2$. This completes the proof.
\vspace{2mm}
\\

We now show that the $E_{\textsc{ab},\nu}^{{\text{\textbf{t}}}_\textsc{a}}$ in equations~\eqref{eq:e1},~\eqref{eq:e2},~\eqref{eq:e3}, and~\eqref{eq:e4} are non-negative. First let us recall that $\fa$, $\fb$, $\theta$ and $\omega$ are real numbers, and that $\fa$ and $\fb$ are both positive. Therefore it is immediately obvious that $E_{\textsc{ab},3}^{{\text{\textbf{t}}}_\textsc{a}}\ge0$.

To show that $E_{\textsc{ab},1}^{{\text{\textbf{t}}}_\textsc{a}}$ is non-negative, note that
\begin{equation}
    -\frac{1}{2}\int\d[n]{\bm k}
    |\ba\pm \bb|^2
    \le 0.
\end{equation}
Exponentiating both sides and expanding the integrand, we find that \mbox{$\exp(\pm\omega)\fa\fb\le 1$}. Therefore
\begin{equation}
    \left(e^\omega+e^{-\omega}\right)
    \fa\fb
    \le 2,
\end{equation}
which implies that $E_{\textsc{ab},1}^{{\text{\textbf{t}}}_\textsc{a}}\ge 0$.

It is a bit more involved to show that $E_{\textsc{ab},2}^{{\text{\textbf{t}}}_\textsc{a}}$ and $E_{\textsc{ab},4}^{{\text{\textbf{t}}}_\textsc{a}}$ are non-negative. We see that $E_{\textsc{ab},2}^{{\text{\textbf{t}}}_\textsc{a}}\ge 0$ if and only if
\begin{equation}
    \left(2-\left(e^\omega +e^{-\omega}\right)\fa\fb\right)^2
    \ge
    4\left|\fa e^{\ii\theta}-\fb e^{-\ii\theta}\right|^2 +
    \left(e^\omega -e^{-\omega}\right)^2 \fa^2\fb^2
    ,
\end{equation}
which is true if and only if $\Gamma_-\ge 0$, where 
\begin{align}
\label{eq:Gamma}
    \Gamma_\pm
    &\coloneqq
    1+\fa^2\fb^2-\fa^2-\fb^2
    \pm\left[\left(e^\omega +e^{-\omega}\right)-
    \left(e^{2\ii\theta}+e^{-2\ii\theta}
    \right)\right]\fa\fb
    \notag\\    
    &=1+\fa^2\fb^2-\fa^2-\fb^2
    \pm 2\left[\cosh(\omega)-\cos(2\theta)\right]\fa\fb.
    \notag
\end{align}
Similarly, we find that $E_{\textsc{ab},4}^{{\text{\textbf{t}}}_\textsc{a}}\ge 0$ if and only if $\Gamma_+\ge0$. Since $\cosh(\omega)\ge 1$ and $\cos(2\theta)\le 1$, we see that $\Gamma_-\le\Gamma_+$, and hence $E_{\textsc{ab},2}^{{\text{\textbf{t}}}_\textsc{a}}\ge 0$ implies $E_{\textsc{ab},4}^{{\text{\textbf{t}}}_\textsc{a}}\ge 0$. Hence to prove our claim that all four of the $E_{\textsc{ab},i}^{{\text{\textbf{t}}}_\textsc{a}}$ are non-negative, it only remains to show that $\Gamma_-\ge 0$.

Let us now define, for simplicity, \mbox{$\bat\coloneqq\sqrt{2}\ba$} and \mbox{$\bbt\coloneqq\sqrt{2}\bb$}, with the $\bnu$ defined in \eqref{eq:bnu}. From the expressions for $\fa$~\eqref{eq:fa_final}, $\fb$~\eqref{eq:fb_final}, $\theta$~\eqref{eq:theta_final}, and $\omega$~\eqref{eq:omega}, we find that
\begin{align}
    \fa&=\exp\left(-\intk |\bat|^2\right),
    \\
    \fb&=\exp\left(-\intk |\bbt|^2\right),
    \\
    2\ii\theta&=-\intk(\bat^*\bbt-\text{H.c.}),
    \\
    \omega&=-\intk(\bat^*\bbt+\text{H.c.}).
\end{align}
From~\eqref{eq:Gamma}, we see that $\Gamma_-$ can now be written as
\begin{align}
    \Gamma_-
    =
    \exp&\left(-\intk(|\bat|^2+|\bbt|^2) \right)
    \notag\\
    \times
    \Bigg[&
    \exp\left(\intk(|\bat|^2+|\bbt|^2) \right)+
    \exp\left(\intk(-|\bat|^2-|\bbt|^2) \right)
    \notag\\
    -&\exp\left(\intk(-|\bat|^2+|\bbt|^2) \right)
    -\exp\left(\intk(|\bat|^2+|\bbt|^2) \right)
    \notag\\
    -&\exp\left(\intk(-\bat\bbt^*-\bat^*\bbt) \right)
    -\exp\left(\intk(\bat\bbt^*+\bat^*\bbt) \right)
    \notag\\
    +&\exp\left(\intk(-\bat\bbt^*+\bat^*\bbt) \right)
    +\exp\left(\intk(\bat\bbt^*-\bat^*\bbt) \right)
    \Bigg].
\end{align}
This can be factored as
\begin{align}
    \Gamma_-
    &=
    \exp\left(-\intk(|\bat|^2+|\bbt|^2) \right)
    \notag\\
    &\hspace{5mm}
    \times\Bigg[
    \Big(
    \exp\big[\intk|\bat|^2 \big]
    -
    \exp\big[-\intk|\bat|^2 \big]
    \Big)
    \Big(
    \exp\big[\intk|\bbt|^2 \big]
    -
    \exp\big[-\intk|\bbt|^2 \big]
    \Big)
    \notag\\
    &\hspace{11mm}
    -\Big(
    \exp\big[\intk\bat\bbt^* \big]
    -
    \exp\big[-\intk\bat\bbt^* \big]
    \Big)
    \notag\\
    &\hspace{11mm}\times
    \Big(
    \exp\big[\intk\bat^*\bbt \big]
    -
    \exp\big[-\intk\bat^*\bbt \big]
    \Big)
    \Bigg]
    \\
    &=
    4\exp\!\left(-\intk(|\bat|^2+|\bbt|^2) \right)\!\!
    \Bigg[\!
    \sinh\!\Big(\intk|\bat|^2\Big)
    \sinh\!\Big(\intk|\bbt|^2\Big)
    -\left|\sinh\!\Big(\intk\bat\bbt^*\Big)\right|^2
    \!\Bigg]\!.
\end{align}
Making use of lemma 1, we find that
\begin{align}
    \Gamma_-
    &\ge
    4\exp\!\left(-\intk(|\bat|^2+|\bbt|^2) \right)\!\!
    \Bigg[\!
    \sinh\!\Big(\intk|\bat|^2\Big)
    \sinh\!\Big(\intk|\bbt|^2\Big)
    -
    \sinh^2\!\Big(\left|\intk \bat\bbt^*\right|\Big)
    \!\Bigg]\!.
\end{align}
Using the fact that $\intk |g(\bm k)|^2\ge |\intk g(\bm k)|^2$ for any function $g(\bm k)$, we obtain
\begin{align}
    \Gamma_-
    &\ge
    4\exp\!\left(-\intk(|\bat|^2+|\bbt|^2) \right)\!\!
    \Bigg[\!
    \sinh\!\Big(\intk|\bat|^2\Big)
    \sinh\!\Big(\intk|\bbt|^2\Big)
    -
    \sinh^2\!\Big(\intk |\bat||\bbt|\Big)
    \!\Bigg]\!.
\end{align}
Applying lemma 2 to the last term, with $g(\bm k)=\bat$ and $h(\bm k)=\bbt$, results in
\begin{align}
    \Gamma_-
    \ge
    4\exp&\left(-\intk(|\bat|^2+|\bbt|^2) \right)
    \notag\\
    &\times
    \Bigg[
    \sinh\Big(\intk|\bat|^2\Big)
    \sinh\Big(\intk|\bbt|^2\Big)
    -
    \sinh^2\Big(\sqrt{\intk |\bat|^2}\sqrt{\intk|\bbt|^2}\Big)
    \Bigg].
\end{align}
The first factor in this expression is clearly positive. From lemma 3, we see that the second factor is non-negative. Hence $\Gamma_-\ge 0$, as desired.

\twocolumngrid

\bibliography{references}
\bibliographystyle{apsrev4-1}

\end{document}